\documentclass[conference]{IEEEtran}
\IEEEoverridecommandlockouts

\usepackage{cite}
\usepackage{amsmath,amssymb,amsfonts}
\usepackage{algorithm}
\usepackage{algorithmic}
\usepackage{graphicx}
\usepackage{textcomp}
\usepackage{xcolor}
\usepackage[hyphens]{url}
\usepackage{enumitem}
\usepackage{wrapfig}

\def\BibTeX{{\rm B\kern-.05em{\sc i\kern-.025em b}\kern-.08em
    T\kern-.1667em\lower.7ex\hbox{E}\kern-.125emX}}
\usepackage{amssymb}
\begin{document}

\pdfpagewidth=8.5in
\pdfpageheight=11in

\newcommand{\iscasubmissionnumber}{NaN}

\pagenumbering{arabic}

\title{Expert Streaming: Accelerating Low-Batch MoE Inference via Multi-chiplet Architecture and Dynamic Expert Trajectory Scheduling}

\author{
    \IEEEauthorblockN{
        Songchen Ma$^{1,2\dagger}$, 
        Hongyi Li$^{2\dagger}$, 
        Weihao Zhang$^{1,2*}$,
        Yonghao Tan$^{1,2}$, 
        Pingcheng Dong$^{1,2}$,  
        Yu Liu$^{1}$,
        Lan Liu$^{3}$,\\
        Yuzhong Jiao$^{1}$,
        Xuejiao Liu$^{1}$,
        Luhong Liang$^{1}$,
        Kwang-Ting Cheng$^{1,2*}$}
    \IEEEauthorblockA{$^{1}$AI Chip Center for Emerging Smart Systems, Hong Kong SAR, China}
    \IEEEauthorblockA{$^{2}$The Hong Kong University of Science and Technology, Hong Kong SAR, China}
    \IEEEauthorblockA{$^{3}$Shanghai UniVista Industrial Software Group Co., Ltd., Shanghai, China}
    \IEEEauthorblockA{$^{\dagger}$ Contributted euqally, $^{*}$ Corresponding Authors (timcheng@ust.hk, weihaozhang@ust.hk)}
}
\maketitle

\maketitle
\thispagestyle{plain}
\pagestyle{plain}


\begin{abstract}
Mixture-of-Experts (MoE) is a promising approach for edge AI with low-batch inference. Yet, on-device deployments often face limited on-chip memory and severe workload imbalance; the prevalent use of offloading further incurs off-chip memory access bottlenecks. Moreover, MoE sparsity and dynamic gating shift distributed strategies toward much finer granularity and introduce runtime scheduling considerations. Recently, high die-to-die (D2D) bandwidth chiplet interconnects create new opportunities for multi-chiplet systems to address workload imbalance and offloading bottlenecks with fine-grained scheduling. In this paper, we propose Fully Sharded Expert Data-parallelism (FSE-DP), a parallelization paradigm specifically architected for low-batch MoE inference on multi-chiplet accelerators. FSE-DP attains adaptive computation–communication overlap and balanced load by orchestrating fine-grained, complementary expert streams along dynamic trajectories across high-bandwidth D2D links. The attendant dataflow complexity is tamed by a minimal, hardware-amenable set of virtualization rules and a lightweight scheduling algorithm. Our approach achieves 1.22-2.00$\times$ speedup over state-of-the-art baselines and saves up to 78.8\% on-chip memory.

\end{abstract}

\section{Introduction}

\begin{figure}[h]
  \centering
  \includegraphics[width=1\linewidth]{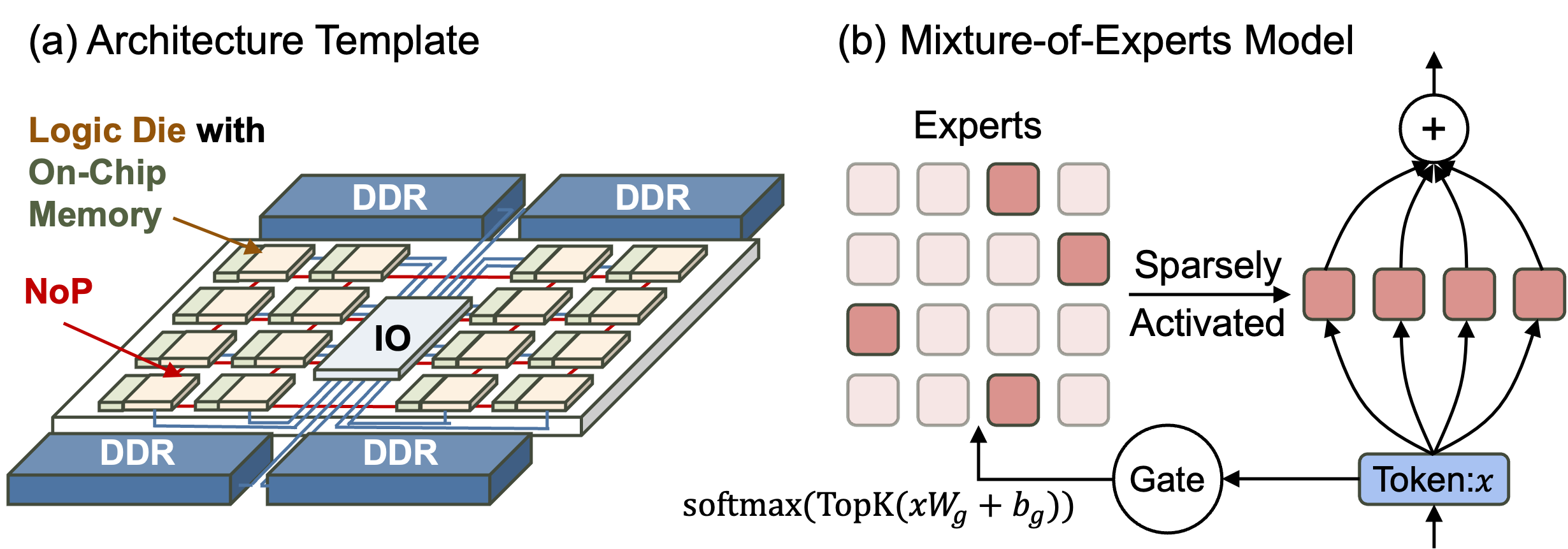}
  \caption{Typical template of (a) multi-chiplet-based AI accelerator. (b) Mixture-of-Experts network.}
  \label{fig:introduction}
\end{figure}

The increasing demand for real-time, privacy-preserving AI services is driving the deployment of Large Language Models (LLMs) onto edge devices. To meet the escalating resource requirements of on-device scenarios such as AI PCs, robotics, and autonomous driving, chiplet-based multi-chiplet accelerators are emerging as a more scalable and cost-effective solution than monolithic designs\cite{zhang2024m2m}. A multi-chiplet package designed for on-device AI typically integrates multiple accelerator dies interconnected by high-bandwidth links and is usually coupled with large off-package memory, such as DRAM\cite{shao2019simba, tan2021nn, lin2024hex} (Figure \ref{fig:introduction}(a)). Concurrently, the Mixture-of-Experts (MoE) architecture has gained prominence for its ability to reduce the number of parameters activated per inference while maintaining a large overall model capacity\cite{fedus2022switch, jiang2024mixtral, liu2024deepseek, yang2025qwen3, wei2024deepseek-ocr, ma20253d} (Figure \ref{fig:introduction}(b)). The combination of multi-chiplet and MoE presents a powerful paradigm for high-performance on-device AI\cite{li2025moe}, where the sparsely activated large model can be distributed across chiplets to leverage spatial parallelism and high-speed inter-die communication.

Despite this promising synergy, effectively deploying MoE models on multi-chiplet packages, particularly in on-device scenarios characterized by low batch sizes, introduces significant challenges. First, \textbf{limited on-chip memory}: Although chiplet technologies significantly enhance the scalability of computational resources, on-chip SRAM-based cache remains a critical asset for on-device systems relative to LLM capacity\cite{srinivasa2025300mb}. Moreover, even devices with large memory dies or GPUs equipped with high-bandwidth HBM still widely adopt off-loading strategies in edge deployments\cite{eliseev2023fast, tang2024hobbit, fang2025klotski, zhang2025daop, cao2025moe}, owing to the persistent demand for ever-larger models. It is therefore essential to minimize redundant storage and maximize memory efficiency\cite{wang2025d, kim2024monde}. Second, \textbf{external memory access bottleneck}: Constrained on-chip capacity forces edge systems to off-load models, necessitating frequent, high-volume off-chip traffic for both KV caches and expert weights. The reduced weight reuse in low-batch scenarios further exacerbates this issue. Third, \textbf{dynamic workload imbalance}: During each forward pass, the number of tokens assigned to each expert varies—some experts process many tokens (hot experts), while others handle few or none (cold experts)\cite{li2023accelerating, kim2024monde, fang2025klotski, zhu2025megascale}. This long-tail distribution introduces two aspects of workload imbalance: (1) the compute-to-data-transfer ratio differs across experts, posing additional obstacles to overlapping off-chip memory access with computation; (2) the storage footprint and compute load diverge among chiplets, reducing overall utilization. Furthermore, low-batch conditions render traditional expert-balancing training\cite{fedus2022switch, wang2024auxiliary} or elastic containers\cite{suo2025coserve, liu2025expert} used in the cloud ineffective. These three challenges are tightly coupled with the characteristics of multi-chiplet architectures, and the latter two are further amplified by low-batch, on-device MoE workloads.

Commonly-adopted parallel strategies—such as expert parallelism (EP)\cite{hwang2023tutel, liu2024deepseek, yang2025hybrid} or hybrids that combine data parallelism (DP)\cite{he2025capacity}, tensor parallelism (TP)\cite{singh2023hybrid, gupta2024lynx, huang2025hd}, and pipeline parallelism (PP)\cite{chen2023pipeline}—fail to adequately address the aforementioned challenges in multi-chiplet inference settings. Most prior optimizations for edge devices target GPUs and seldom exploit the distinctive characteristics of chiplet-based packages. Recently, several methods specialized for MoE inference on multi-chiplet have been proposed\cite{he2025hydra, yu2025orders}. These approaches generally aim to optimize inter-die communication, especially all-to-all communications in MoE, but they place less emphasis on addressing the long-tail issue and the external memory access bottleneck.

Propelled by rapid advancements in advanced packaging and high-speed communication technologies, the emergence of cost-effective and power-efficient die-to-die (D2D) interconnects with massive bandwidth and low latency—now being standardized by protocols like UCIe~\cite{sharma2022universal}—is fundamentally recasting inter-die data transfer from a performance bottleneck into a rich architectural resource. \textbf{Capitalizing on this significant opportunity, a novel parallelization strategy is expected to not only fully exploit the intra-package communication resource to enhance on-chip memory efficiency and curtail off-chip traffic, but also replace the expensive collective communication, inherent to traditional EP/TP, with highly efficient point-to-point transfers.} Complementing the interconnection benefit, each chiplet typically possesses an independent control path, enabling chiplet arrays to exhibit a Multi-Instruction-Multi-Data (MIMD) character that supports asymmetric execution patterns. Consequently, load balance can be systematically engineered through a non-uniform yet mutually complementary mapping space\cite{shan2022architecture,zhang2024m2m}. Collectively, these two factors unlock opportunities for fine-grained, synchronization-free dataflow among chiplets\cite{zhang2023indm}. Grounded in these observations, we propose Fully Sharded Expert Data-parallelism (FSE-DP), a specialized parallelization strategy tailored for MoE inference on multi-chiplet packages that delivers the following advantages:

\textbf{Save on-chip memory and reduce off-chip memory traffic.} The root cause of duplicated on-chip memory is the rigid "one-chip-one-slice" mapping assumed by EP or TP. FSE-DP breaks this assumption by treating the whole chiplet array as a single, pooled buffer: only one physical copy of any token/expert slice is kept in the entire package. Leveraging high D2D bandwidth, FSE-DP \emph{streams} the expert slice along a scheduled trajectory. The saved on-chip memory provides more data-reuse opportunities to reduce external memory access.

\textbf{Dynamic computation-communication overlap and load balancing.} The long-tail distribution and dynamic features of MoE render fixed scheduling infeasible. FSE-DP introduces a dynamic fine-grained dataflow to realize each expert's trajectory, which turns this heterogeneity into an opportunity. By fusing and complementing the dataflow of expert trajectories with different load characteristics at fine granularity, FSE-DP achieves dynamic computation-communication overlap under non-unified memory access (D2D and die-to-DDR), minimizes on-chip memory, and enables load balancing.

\textbf{Streamline fine-grained complexities with hardware-efficient rules and algorithms.} The fine-grained dataflow fusion ostensibly introduces intricate memory-access and communication patterns, significantly elevating execution complexity. Nevertheless, FSE-DP's dataflow is steered by a handful of lightweight, self-acting rules. Each chiplet receives the slice, computes its local token batch, and immediately forwards the slice to the next chiplet. These rules transparently abstract away memory and inter-die communication details from the programmer while letting the hardware spontaneously materialize the expert trajectories. Building upon this virtualization, a hardware-efficient runtime-scheduling algorithm and a dedicated hardware scheduler can be devised.

Overall, this paper makes the following contributions:

\begin{itemize}
  \item Leveraging high D2D bandwidth, we introduce FSE-DP, a distributed-parallel strategy tailored for multi-chiplet architectures that eliminates on-chip redundancy and reduces off-chip DRAM traffic.

  \item Architecting a fine-grained dataflow for expert trajectories and \textcolor{black}{fusing heterogeneous flows} to achieve adaptive computation-communication overlap and load balancing. Along with this, we present a paired-load policy and token-buffering policy to mitigate bandwidth bottlenecks.

  \item Establishing a minimal set of virtualization rules \textcolor{black}{for execution abstraction} that automatically orchestrate dynamic expert trajectories under diverse workload scenarios, drastically simplifying both software programming and hardware-runtime complexity.

  \item Presenting a hardware-efficient scheduling algorithm that unites temporal QoS-pressure-based queuing with spatial expert-trajectory planning. 

  \item {Developing a system to accelerate MoE-based large model inference. \textcolor{black}{Our system includes a taped-out $2\times2$ 5nm MCM test chip integrating a UCIe-compliant high-speed D2D interconnect, along with a lightweight, specialized scheduler implemented as a synthesized RTL module that realizes our proposed algorithm.}}
  
  \item {Comprehensive evaluations on our system demonstrate that our approach not only achieves significant performance gains over state-of-the-art baselines but exhibits scalability and robustness.}
\end{itemize}

\section{Background and Motivation}
\label{sec:background}

\subsection{Related Works}
MoE mitigates the widening gap between exploding model capacity and constrained hardware by activating only a sparse subset of expert sub-networks for each token. A gating function selects the Top-K experts and aggregates their outputs, an idea that can be applied to any parameter block, including the attention layer (Mix-of-transformers, MoT)\cite{wang2024moa, jin2024moh}. Originally popularized in the cloud, relatively small-scale MoE is now rapidly developed for edge scenarios\cite{shen2024jetmoe, qwen1.5, yang2025qwen3, wu2024yuan, zhu2024llama, abouelenin2025phi, liu2024grin, muennighoff2024olmoe, liu2025muon}.

Current MoE optimizations primarily focus on GPU systems. Cloud-scale MoE deployments on GPU clusters predominantly optimize two points: (1) the all-to-all token–expert permutation traffic inherent in EP, and (2) load imbalance caused by skewed expert popularity. Hybrid parallelism (EP+TP+DP)\cite{hwang2023tutel, singh2023hybrid, huang2025hd}, pipelined or fused collective communication\cite{shi2023pipemoe, punniyamurthy2024optimizing, wang2025harnessing}, and specialized communication libraries\cite{rajbhandari2022deepspeed, liu2024deepseek} are the main research topics for reducing inter-node traffic. For load balancing, auxiliary-loss-based training\cite{fedus2022switch, wang2024auxiliary} or elastic expert computation\cite{yu2024moesys, doucet2025harmoeny} are common strategies to keep workloads uniform under large batches. \textcolor{black}{MoETuner\cite{go2025moetuner} optimizes expert placement across GPUs by solving an integer linear programming that jointly considers per-expert token load and inter-layer routing dependencies, reducing inter-GPU token routing skew and tail latency.} Some techniques use Fully Sharded Data Parallelism (FSDP)\cite{zhao2023pytorch} to optimize MoE training, further sharding individual experts across GPUs and replacing all-to-all with cheaper all-gather/reduce-scatter operations\cite{pan2025fsmoe}.

\textcolor{black}{Rotation-style distributed GEMM dataflows\cite{gao2019tangram, he2025waferllm} also explore how structured cyclic shifting can improve locality and overlap when data movement is unavoidable. For example, WaferLLM's MeshGEMM targets wafer-scale mesh NoCs and accelerates the prefill phase by combining cyclic shifting with an interleaving mapping, so that each core exchanges tiles with a fixed set of nearby neighbors and bounds the per-step communication cost to a constant hop distance (reducing both long-range latency and routing pressure). More broadly, these designs suggest that converting global exchange into neighbor transfers can help overlap data movement and computation, \textcolor{black}{although they primarily target \emph{static}, \emph{dense}, and \emph{predictable} GEMM.} These efforts provide insights that we extend to chiplet systems via pure point-to-point weight transfers.}

For on-device inference scenarios with very limited GPU memory (e.g. NVIDIA RTX 3060 laptop only has 6GB memory), systems usually rely on off-loading strategies, paging experts to CPU memory or SSD. This creates a cross-level data exchange and turns external bandwidth into the dominant bottleneck. Consequently, on-device-oriented studies focus on (1) expert prefetching with learned or heuristic predictors\cite{eliseev2023fast, du2024sida, yao2024exploiting, song2024promoe, li2025speculative}, (2) on-chip caching of hot experts\cite{he2024expertflow, zhong2024adapmoe}, and (3) run-time schedulers that overlap expert I/O with computation\cite{kossmann2022optimizing, li2025static, wang2025d}.

Emerging multi-chiplet accelerators open an under-explored design space. Current studies mainly focus on expert mapping/placement to mitigate all-to-all communication\cite{he2025hydra, yu2025orders} or propose Content-Addressable Memory (CAM) to bypass token permutation\cite{he2025hydra}; some also explore run-time migration of experts toward near-memory processors\cite{huang2025hd}. Yet, in on-device multi-chiplet settings, the off-loading pressure inherent to GPU-based systems and the statistical load imbalance persist simultaneously. 

\subsection{Motivation}

\begin{figure}[h]
  \centering
  \includegraphics[width=0.9\linewidth]{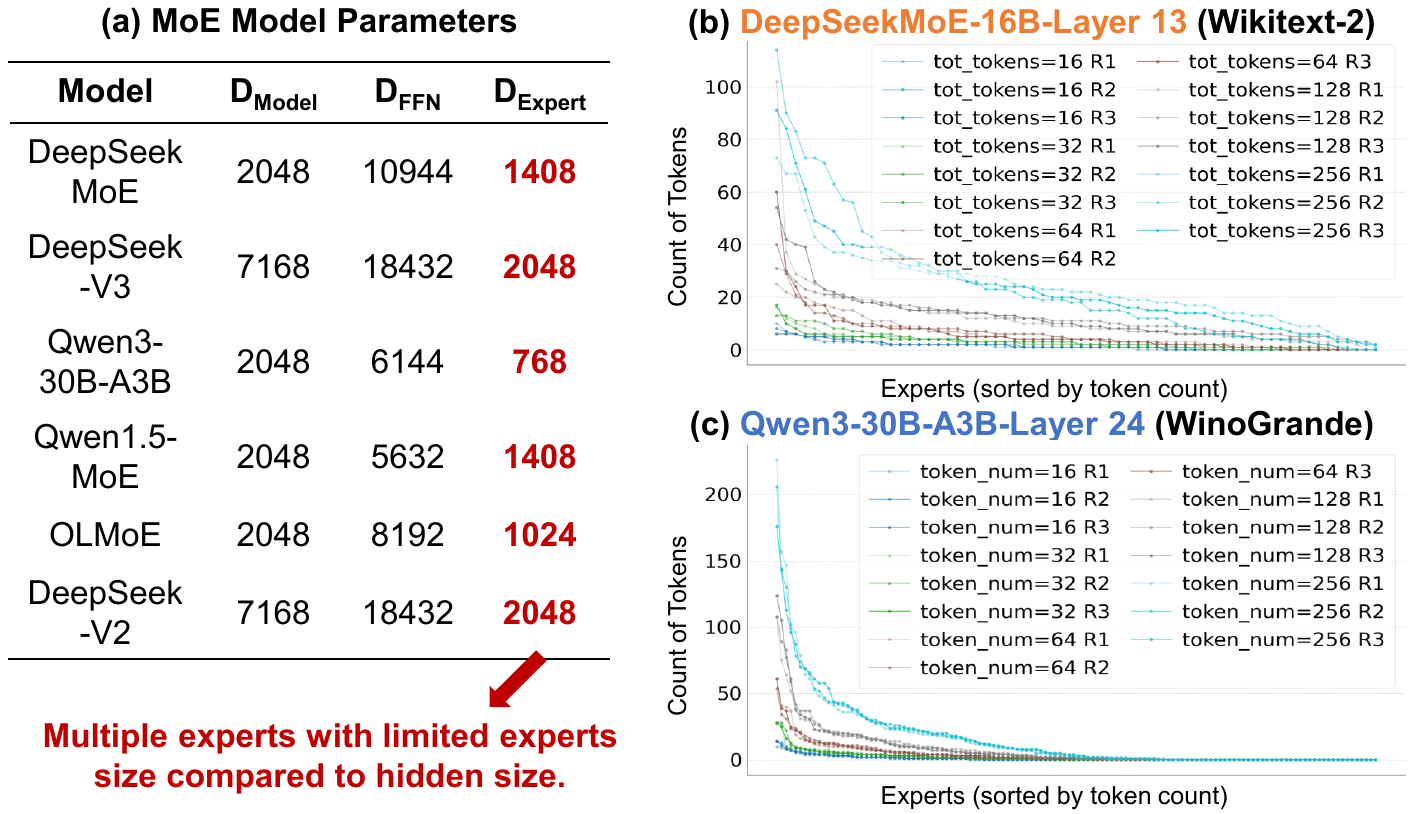}
  \caption{\textcolor{black}{(a) Shapes for different models (b,c) Long-tail effect of MoE models under different batch sizes. The figure shows the number of tokens processed in a specific layer for DeepSeek-MoE-16B\cite{rajbhandari2022deepspeed} on the Wikitext-2 dataset\cite{merity2016pointer} and Qwen3-30B-A3B\cite{yang2025qwen3} on WinoGrande\cite{sakaguchi2021winogrande}. Experts on the x-axis are sorted by the number of tokens they process; the y-axis gives the token count per expert. The long-tail effect is more pronounced at smaller token numbers. R denotes different requests.}}
  \label{fig:observation}
\end{figure}

Figure \ref{fig:observation}(a) characterizes contemporary MoE models. The results show that the dimension of a single expert ($D_{Expert}$) is generally smaller compared to the FFN ($D_{FFN}$) and hidden size ($D_{Model}$). This makes the computational granularity of the experts \textcolor{black}{finer-grained, with lower computational demands} but still places very high demands on memory bandwidth\cite{zhu2025megascale}. 

Furthermore, we profile expert activation in state-of-the-art MoE networks on \textcolor{black}{language datasets}. Figure \ref{fig:observation} (b) and (c) highlight the pronounced \textbf{long-tail effect}: although these MoE models incorporate expert-balancing losses during training, inference still exhibits large disparities in per-expert token counts for batched tokens ranging from 16 to 256. A non-negligible fraction of experts process only a handful of tokens, yet their full weights must be fetched, creating severe bandwidth pressure and necessitating strategies that maximize data reuse once loaded onto the chip. However, existing DP and sequence parallelism (SP)\cite{li2021sequence} replicate entire experts across chiplets, incurring weight redundancy. TP shards experts but duplicates tokens, while EP also mandates token replication, compounding the memory-traffic challenge.

\textcolor{black}{Many edge deployments also operate in a \emph{low-batch} regime that aggregates the long-tail effect: on-device systems increasingly run multiple concurrent, mixed-latency tasks (e.g., agentic workloads with time-varying concurrency, robotics pipelines that couple perception/reasoning/action, and multi-sensor streaming). Compared with cloud serving, effective concurrency is typically smaller, which weakens cross-request weight reuse and makes off-chip expert fetch under long-tailed activations a primary bottleneck.}

The evolution towards multi-chiplet AI accelerators, driven by advanced packaging, has unlocked new architectural affordances for large-scale models: ultra-high-bandwidth, low-latency D2D links and per-chiplet independent control that enables MIMD-style asymmetric dataflows. \textcolor{black}{Multi-chiplet design disaggregates the system into smaller, function-specific dies for better manufacturing yield and design reuse\cite{ChipletDesignRule}, allowing for fine-grained, synchronization-free pipelines rather than device/rack-level pipelines in GPU-based systems.} This progression began with standard packaging, Multi-Chip-Modules (MCMs), exemplified by Nvidia's Simba accelerator, which used a proprietary D2D interconnect to achieve 100~GB/s/chiplet bandwidth at 20~ns/hop and 0.82–1.75~pJ/bit~\cite{shao2019simba}. Subsequent advancements in 2.5D heterogeneous integration~\cite{chou2022netflex,tan2023scalable} have pushed performance further, with recent systems demonstrating an aggregate bandwidth of 20~Tb/s across 20 chiplets~\cite{srinivasa2025300mb}. However, realizing the full potential of this packaging spectrum—from cost-effective 2D MCMs~\cite{zhu2022comb, tu202316} to high-performance 2.5D systems—was historically hindered by a fragmented landscape of proprietary interconnects (e.g., AMD's Infinity Fabric) and early open specifications (e.g., OCP-ODSA's BOW). The emergence of the UCIe standard~\cite{sharma2022universal} marked a pivotal shift, creating a unified ecosystem by providing optimized options for both advanced (UCIe-A) and standard (UCIe-S) packages~\cite{mota2023ucie}. This dual-pronged strategy's strength is confirmed by recent industrial implementations. For advanced packages, D2D links in 3nm have demonstrated a remarkable bandwidth density of up to 10.5 Tb/s/mm \cite{lin202536}, while other work has shown an energy efficiency of just 0.29 pJ/b \cite{melek20250}. Concurrently, for standard packages, 3nm UCIe-S transceivers provide a competitive 0.448 Tbps/mm at 0.52 pJ/bit~\cite{vandersand20250}. This convergence of versatile packaging technologies with a proven, unified open standard~\cite{srinivasa2025300mb,tan2023scalable,jiao202537} is finally delivering the low-latency, energy-efficient, and high-bandwidth D2D interconnects essential for next-generation computing. This robust hardware foundation creates an unprecedented opportunity to leverage a rich ``communication bandwidth resource," yet exploiting its diverse affordances demands novel parallelization strategies.

\section{Naive Fully Sharded Expert Parallel}
\label{sec:basic_fsedp}

\begin{figure*}[h]
  \centering
  \includegraphics[width=0.92\linewidth]{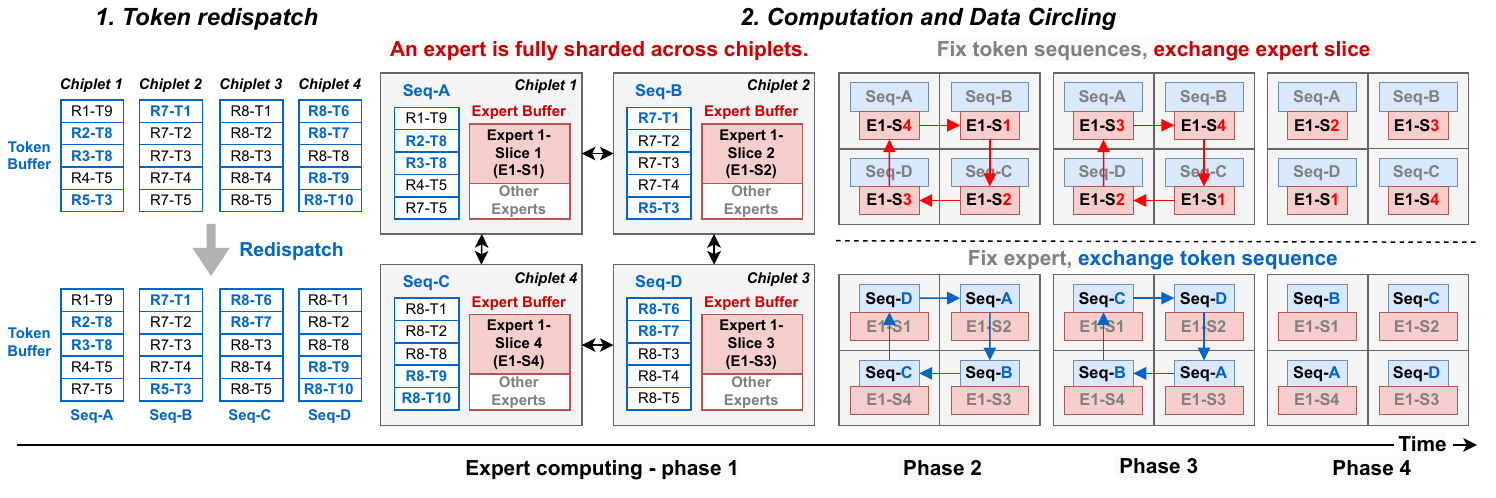}
  \caption{\textcolor{black}{\textbf{Fully sharded expert–data parallelism (example: four chiplets compute expert 1).} The figure illustrates how a single expert is computed within one MCM: expert 1 is evenly sharded into slices across chiplets (E1-S1--E1-S4). black tokens denote the tokens that activate expert 1, while non-highlighted (gray/black) tokens are other buffered tokens on that chiplet that do not activate expert 1. Before computing expert 1, tokens are \emph{redispatched} across chiplets to balance the number of black tokens per chiplet for load balancing. $R$ denotes a request, and Seq denotes the activated token sequence stored on a chiplet.} During expert computation, chiplets can exchange data in two equivalent ways to cover all black tokens: (a) keep token sequences fixed and circulate expert-1 slices so each slice visits the chiplets holding black tokens; or (b) keep expert 1 slices fixed and circulate black-token sequences so they visit all chiplets containing slices of expert 1.}
  \label{fig:fsedp}
\end{figure*}

When the D2D interconnect ceases to be a performance bottleneck and instead becomes an exploitable resource, the optimization paradigm shifts from intra-die communication, which is the main focus of existing parallel strategies, to off-chip access and the elimination of on-chip redundancy, thereby maximizing data reuse. Building on this insight, we introduce Fully Sharded Expert-Data Parallelism (FSE-DP). Inspired by FSDP in distributed neural-network training, where both inputs and model parameters are sharded across devices, FSE-DP also partitions token sequences and expert weights across chiplets.

We refer to each forward pass as an iteration. During an iteration, FSE-DP processes the MoE network layer-wise, keeping token activations on-chip while fetching expert weights from DDR on demand. To illustrate the core idea, consider a 4-chiplet array that handles one expert at a time (Figure~\ref{fig:fsedp}). The current iteration aggregates tokens from multiple requests (combining prefilling and decoding, a widely adopted strategy called chunked prefill~\cite{agrawal2023sarathi}). These tokens are evenly sharded across the four chiplets, so every chiplet holds one fourth of the tokens. In Figure~\ref{fig:fsedp}, ``R1-T9'' denotes the activation vector of the 9th token from Request-1. For each expert, the set of tokens that activate it varies dynamically. Consequently, prior to computing each expert, we redistribute the tokens that activate the current expert to ensure that the number of tokens processed by each chiplet is approximately equal for load balancing. In this example, we designate the token sequences computed by each chiplet as Seq-A to Seq-D, with the lengths of Seq-A through Seq-D kept roughly equal.

We also partition expert 1 into four slices (E1-S1 through E1-S4) across chiplets. In the first computation phase, illustrated in Figure \ref{fig:fsedp}, chiplet 1 processes Seq-A with E1-S1, chiplet 2 processes Seq-B with E1-S2, and so on. In the second phase, as shown in Figure \ref{fig:fsedp}(a), we perform a circular transfer of expert slices across chiplets: chiplet 1 sends E1-S1 to chiplet 2; chiplet 2 sends E1-S2 to chiplet 3; chiplet 3 sends E1-S3 to chiplet 4; and chiplet 4 sends E1-S4 to chiplet 1. Each chiplet then computes its token sequence using the newly received expert slice. In subsequent phases, this pattern of expert-slice circulation and computation continues under the same rules until all computations for this expert are completed.

In current MoE-parallelism approaches, transmitting expert weights is seldom adopted because the weight volume typically exceeds that of token activations. Indeed, when a multi-chiplet system computes a single isolated expert, fixing the weights and exchanging token sequences achieves comparable parallelism, as illustrated in Figure \ref{fig:fsedp}(b). When the token sequence is small, inter-chiplet token transmission is more attractive. However, as discussed earlier, properly engineered advanced interconnects broaden the design space for weight-transmission strategies. When D2D communication is acceptable, transmitting expert slices offers additional benefits, enabling improved overlap between DDR loading and computation while reducing on-chip memory and bypassing token redistribution.

\section{FSE-DP with Micro-Slice Flow}
\label{sec:fsedp_flow}

\begin{figure*}[h]
  \centering
  \includegraphics[width=0.92\linewidth]{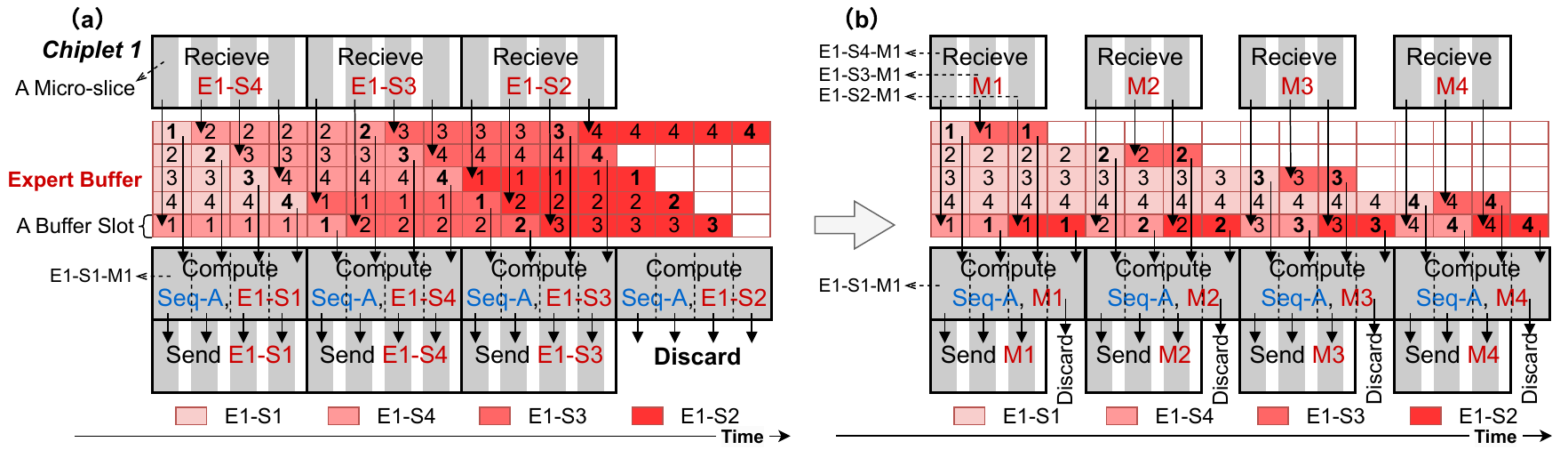}
  \caption{\textcolor{black}{\textbf{Micro-slice flow for overlapping D2D communication and computation (example: chiplet 1 while computing expert 1).} Each expert slice (E1-S1--E1-S4) is further partitioned into micro-slices (M1--M4). \textbf{(a) Baseline micro-slice overlap.} In each step, chiplet 1 computes its local sequence using the current micro-slice, while concurrently receiving the next micro-slice from a neighbor chiplet and sending the just-computed micro-slice to the next chiplet; arrows indicate the D2D transfers that overlap with the compute stage. The weight buffer show the micro-slice storage on chiplet 1 over time, where each row is one micro-slice-sized buffer slot and each column is a time step. A colored cell indicates that the slot stores a micro-slice from a specific expert slice (see the color legend), and the numeral in the cell is the micro-slice index within that slice. Blank cells are free slots, and bold numerals mark the micro-slice being computed in that step. \textbf{(b) Eager micro-slice usage.} Chiplet 1 immediately forwards the micro-slice under computation and, in the next step, computes the most recently received micro-slice, so each micro-slice quickly traverses all chiplets and can be discarded earlier, reducing average weight-buffer occupancy.}}
  \label{fig:micro-slice}
\end{figure*}

We improve FSE-DP by centering on computation-communication overlap. Computation-communication overlap is essential in neural-network acceleration to mitigate performance loss from communication overhead. In multi-chiplet systems, beyond the attention phase, overlap should be exploited in three key scenarios: (1) inter-chiplet data exchange, where chiplets compute on the current data while simultaneously transmitting the next expert slice or token sequence; (2) on-chip computation of the current expert while pre-loading weights for the next expert from DDR. Considering these scenarios, we identify two principal limitations of the basic FSE-DP approach: (1) regardless of whether weights or tokens are exchanged between chiplets during expert computation, each chiplet requires extra storage to hold data for the upcoming computation; for example, when expert slices are transmitted in FSE-DP, every chiplet needs additional space for the next expert slice, nearly doubling the on-chip memory requirement for expert weights; and (2) although FSE-DP alone can balance computation and storage, coordinating it with attention computation is challenging because varying KV-cache sizes across requests preclude uniform partitioning of expert and sequence slices at the start of expert computation.

To reduce the storage overhead of communication buffers, a straightforward approach is to reduce the granularity of computation-communication operations. Using expert weight transmission in FSE-DP as the example, we further partition each expert slice on a chiplet into multiple micro-slices, treating each micro-slice as the fundamental unit for FSE-DP computation and transmission. As illustrated in Figure \ref{fig:micro-slice}(a), while a chiplet computes one micro-slice of its current expert slice, it concurrently receives a micro-slice from the next expert slice scheduled for computation. Upon completing the current micro-slice, the chiplet releases the associated storage space. After all micro-slices of the current expert have been processed, the chiplet will have accumulated the full set of weight slices for the next expert. This approach can be implemented with a micro-slice–based ring buffer—a mature hardware technique—thereby reducing the additional storage overhead of computation-communication overlap to the size of a single micro-slice. \textcolor{black}{The number of micro-slices reflects a trade-off between overlap opportunity and overhead. Finer micro-slices improve pipelining and reduce the required communication-buffer, but also increase relative control/scheduling overhead and per-transfer header cost, leading to diminishing returns when a micro-slice becomes too small. A practical approach is to choose a micro-slice size such that its computation time is roughly comparable to its D2D transfer time, maximizing overlap while preventing control/dispatch overhead from becoming the bottleneck.}

Departing from the straightforward approach, we introduce an unconventional micro-slice–centric optimization of the fine-grained flow shown in Figure \ref{fig:micro-slice}(a). When a token completes computation in a given layer, it produces a new activation vector of the same size. By contrast, once weight computation finishes, those weights are not reused within the current iteration; hence, their on-chip memory can be released immediately. Guided by this observation, the principle for fine-grained optimization is to \textbf{complete all computations for each micro-slice as quickly as possible, release its space, and then proceed with computations for other micro-slices}. Applying this rule yields the pipelined pattern shown in Figure \ref{fig:micro-slice}(b). Although the pattern may appear complex, its rules are simple: \textbf{each chiplet immediately transmits the micro-slice it is currently computing and, in the next time step, processes the most recently received micro-slice}. These rules ensure that once a micro-slice begins computation, it is promptly swept into the dataflow circulating among chiplets, completes all token computations, and occupies on-chip memory for the shortest possible time. As shown in Figure \ref{fig:micro-slice}(b), this approach can save nearly half of the on-chip memory (more than half if an expert is not pre-loaded).

\subsection{Expert Load Optimization}
\label{sec:expert_load}

\begin{figure}[h]
  \centering
  \includegraphics[width=0.9\linewidth]{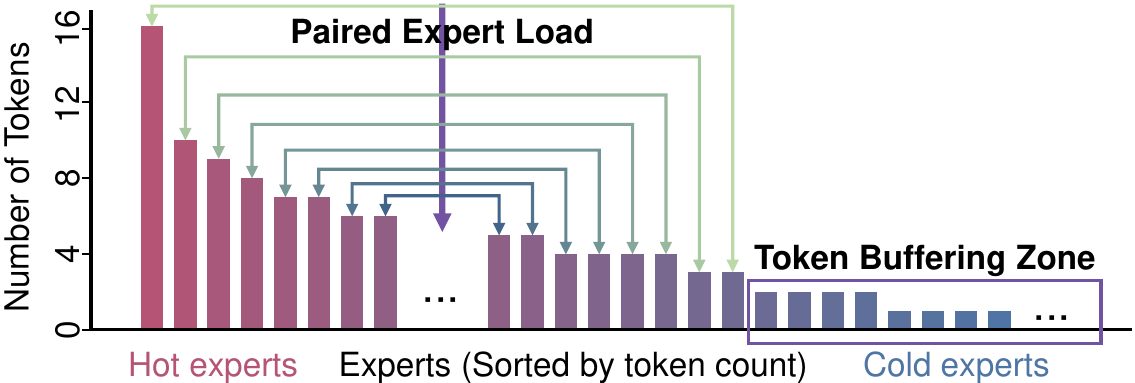}
  \caption{Demonstration of paired-load policy and token buffering policy.}
  \label{fig:expert_laod}
\end{figure}

To capitalize on the buffer headroom and timing regularity created by the micro-slice flow and to prepare the subsequent fusion of DDR load with D2D flow, we first present our expert-load order optimization. Hot experts incur intensive computation workloads, whereas cold experts exhibit prominent communication bottlenecks in the long tail, which suggests pairing hot and cold experts. Accordingly, we sort experts by their token-activation counts and pair experts from opposite ends of the list, interleaving on-chip fetch and compute, as shown in Figure \ref{fig:expert_laod}. This \textbf{paired-load policy} aligns with the micro-slice cadence, increases overlap between communication-bound and compute-bound experts (Section \ref{sec:flow_fusion}).

For long-tail experts activated by only a few tokens, loading them onto the chip leads to highly inefficient bandwidth utilization due to low data reuse. \textcolor{black}{Accordingly, rather than processing these tokens in the current iteration, we use \textbf{token buffering} as a \emph{per-request deferral} mechanism at the specific MoE layer: when a request's tokens are routed to an extremely cold expert that is not scheduled for immediate execution, the scheduler can pause the \emph{entire request} at that MoE layer, holding its intermediate activations. In the following iterations, these tokens can be combined with newly arriving tokens of other requests to evaluate expert-activation patterns.} While this increases that request's latency, LLMs typically require multiple forward passes during decoding. For example, when generating 4k tokens of text, permitting a 10\% increase in total completion time yields more than 400 opportunities for token buffering. Moreover, in LLM scenarios, per-request quality-of-service (QoS) requirements are often flexible, making it reasonable to trade some per-request performance for improved overall system efficiency. In Section \ref{sec:schedule}, we present detailed methods for applying token buffering.

\textcolor{black}{The paired-load and token-buffering policies are compatible with expert-prediction prefetch techniques, such as Pre-Gated MoE~\cite{hwang2024pre}. During the computation between the previous layer’s FFN and the current layer’s attention, expert load can be reordered based on predicted expert-activation patterns. With accurate prediction, the scheduler can commit to an expert order early and start prefetching to better hide latency. If predictions are inaccurate, both load reordering and token buffering can be updated again after the MoE gate of the current layer. Moreover, because scheduling operates at expert granularity, much of the remaining decision latency can overlap with the execution of other experts.}

\subsection{Flow Fusion}
\label{sec:flow_fusion}

\begin{figure}[h]
  \centering
  \includegraphics[width=1\linewidth]{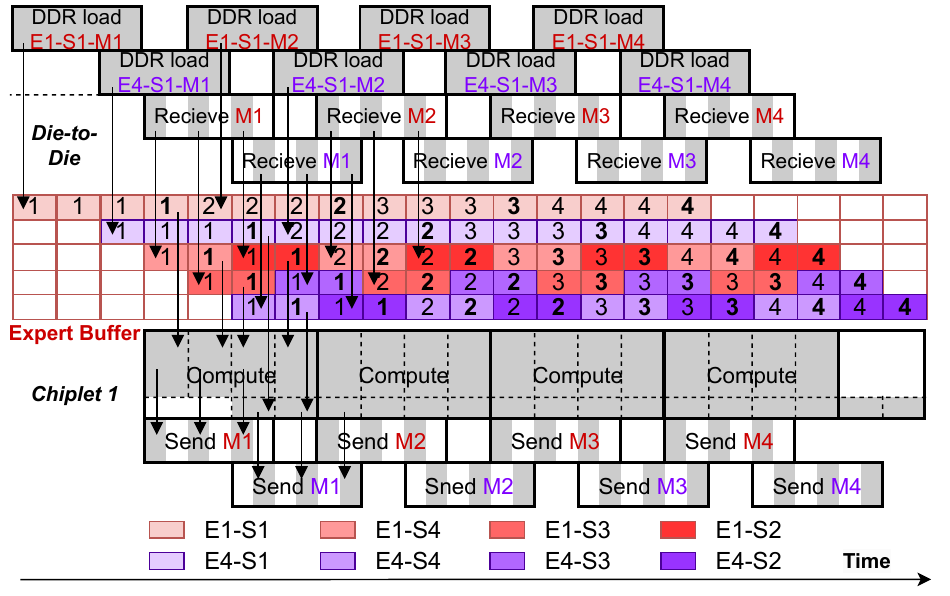}
  \caption{\textcolor{black}{\textbf{Flow fusion of DDR load and D2D micro-slice flow (example: chiplet 1 while computing expert 1 paired expert 4).} This figure extends Figure~\ref{fig:micro-slice}(b) by fusing off-chip DDR loading with the on-chip D2D micro-slice circulation, illustrated on chiplet 1 while co-executing two experts (E1 in red and E4 in purple under paired-load). DDR loads of the locally assigned micro-slices (e.g., E1-S1-M* and E4-S1-M*) are pipelined with computation and D2D receive/send of micro-slices from other chiplets. DDR load latency is assumed to be $4\times$ one micro-slice compute step. The expert-buffer grid is read as in Figure~\ref{fig:micro-slice}: the first two rows cache DDR-loaded micro-slices for E1 and E4, respectively, while the remaining rows serve as a shared staging area for interleaved D2D-received micro-slices from both experts. Bold numerals indicate the micro-slices being computed at each time step.}}
  \label{fig:micro-slice-ddr}
\end{figure}

While the micro-slice flow effectively optimizes D2D communication, the primary performance bottleneck often remains the communication between the chip and off-chip DDR memory, which is substantially slower. Accordingly, we leverage the on-chip memory saved by the D2D flow to fuse it with the chip-to-DDR flow, thereby achieving further overlap of computation with DDR loading within the same on-chip memory budget. We illustrate this fusion using chiplet 1 in Figure \ref{fig:micro-slice-ddr}. In this example, we complementarily load and compute E1 and E4 on-chip (paired-load policy): chiplet 1 loads slice 1 of E1 and E4 from DDR and receives slices 2–4 from other chiplets. We assume that loading a micro-slice from DDR takes four times as long as computing a micro-slice.

In Figure \ref{fig:micro-slice-ddr}, the first two rows of the weight buffer store micro-slices of E1 and E4 loaded from DDR, respectively, while the subsequent three rows are reserved for the interleaved reception of micro-slices from other chiplets. Through fine-grained fusion of the DDR-load flow and the D2D flow, we overlap transmission and computation for two experts. At each time step, a chiplet computes one micro-slice from E1 and one from E4, preserving load balance while keeping the maximum expert-buffer usage the same as in Figure \ref{fig:micro-slice}(b). This figure shows an ideal model to illustrate the principle. Note that different ratios of DDR load time to micro-slice compute time yield different expert-buffer utilization patterns. Ratios larger than those shown in Figure \ref{fig:micro-slice-ddr} may require more buffer storage (Section \ref{sec:dse}), although the allocation rules and overall fusion pattern remain consistent (Section \ref{sec:virtualization}).

\subsection{\textcolor{black}{Execution Abstraction via Virtualization Rules}}
\label{sec:virtualization}

Up to this point, we have detailed the benefits of fine-grained pipelining and its basic arrangement. Nevertheless, a significant challenge arises: under dynamic and imbalanced workloads, when multiple experts are fused concurrently, and with heterogeneous DDR and D2D links, the resulting micro-slice storage and communication patterns can become exceedingly complex. Even the flow in Figure \ref{fig:micro-slice}(b)—an idealized case with two uniformly distributed experts—already exhibits substantial complexity for manual pipeline scheduling. \textbf{Directly orchestrating this level of complexity at runtime is prohibitive for both software programmability and hardware implementation. Therefore, this section introduces a ``virtualization'' method.} \textcolor{black}{The ``virtualization'' abstracts away dynamic physical details, e.g., which micro-slice resides in which buffer slot and which link it traverses at which cycle. Behind a small set of rules, the scheduler can reason at the level of expert trajectories rather than per-micro-slice bookkeeping. By following these rules, complex flow orchestration can be realized while ensuring low hardware implementation costs.}

From our previous discussion, the core of FSE-DP is to enable expert parameters to flow through all chiplets responsible for token computation along a trajectory. As long as each micro-slice visits every station, the starting point, endpoint, timing, and intra-trajectory ordering of micro-slices are immaterial (the result of a token–expert computation can be accumulated on the same chiplet without tracking which micro-slice is being processed). Consequently, the specific storage locations of micro-slices or token activations on chiplets, and their transmission schedules among chiplets, are of secondary importance, provided their paths follow the trajectory. \textcolor{black}{In our dynamic scheduling, trajectories are decided at expert granularity: different experts may choose different chiplets and flow directions at runtime, while once a trajectory is selected for an expert in a scheduling iteration, all micro-slices of that expert follow the same trajectory. We do not implement per-micro-slice dynamic paths because tracking per-micro-slice trajectory state and next-hop decisions would substantially increase scheduler metadata and hardware complexity. }

\begin{figure}[h]
  \centering
  \includegraphics[width=1\linewidth]{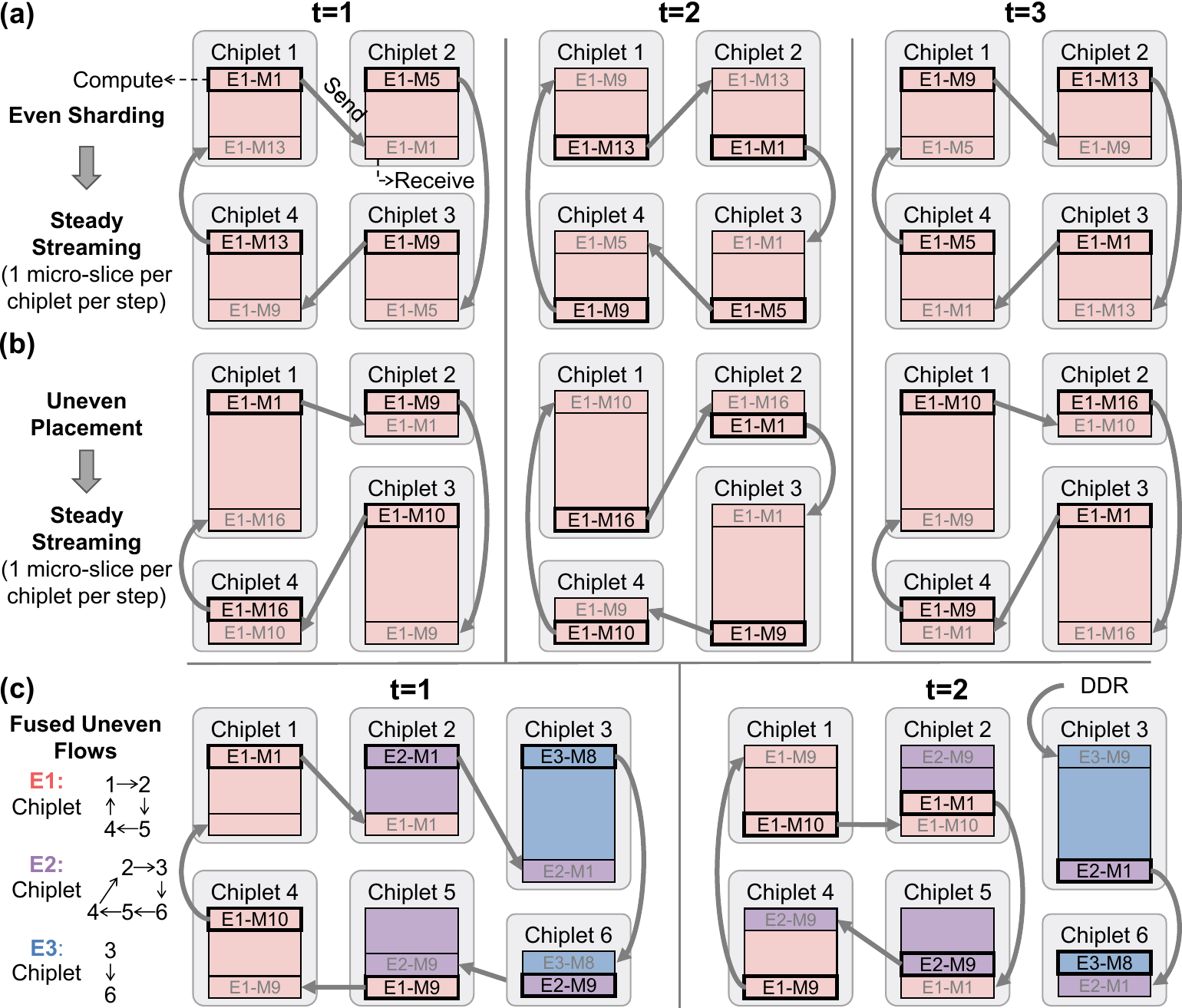}
  \caption{\textcolor{black}{\textbf{Virtualization demonstration under micro-slice streaming (example: 4 chiplets, ring expert trajectory).} Each panel shows the on-chip storage state of the four chiplets at three consecutive time steps ($t{=}1$--$3$): each block is a stored micro-slice (labeled by expert and index, e.g., E1-M1), and arrows indicate next-hop forwarding along the ring trajectory. (a) Baseline case where expert 1 is evenly sharded across chiplets; achieving efficient overlap with explicit buffer assignment requires careful placement (as in Figure~\ref{fig:micro-slice-ddr}). (b) Expert~1 is unevenly distributed across chiplets, yet the computation proceeds correctly once micro-slices start streaming along the trajectory. For simplicity, (a) and (b) assume a ring topology in which micro-slices move from Chiplet~1 to~2, 2 to~3, and so on, to introduce the background of virtualization.(c) There experts execute concurrently with highly uneven on-chip arrangement. Despite the seemingly irregular trajectories, virtualization rules abstract away these runtime details and still realize a smooth flow. The key message is that the hardware scheduler need not reason about per-chiplet implementation: as long as micro-slices follow the prescribed trajectory, correct and efficient execution emerges without fine-grained bookkeeping.}}
  \label{fig:virtualization}
\end{figure}

\textcolor{black}{Figure \ref{fig:virtualization} illustrates how runtime storage details are irrelevant to the smooth streaming of micro-slices along an expert trajectory. Figure \ref{fig:virtualization}(a) shows the baseline case where expert 1 is evenly sharded across 4 chiplets. The per-chiplet weight buffer is shown over three time steps. Figure \ref{fig:virtualization}(b) then shows that even when E1's micro-slices are unevenly distributed across chiplets, once the micro-slices begin to flow among chiplets, we can still realize a fine-grained flow equivalent to that in Figure \ref{fig:micro-slice}(b) while maintaining balanced expert computation across all chiplets. \textbf{For clarity, we simplify the notation by moving from a two-level partitioning (expert slice to micro-slice) to a single level, in which an expert is directly partitioned into a set of micro-slices.} For instance, if expert E1 is divided into 16 micro-slices, we denote them as E1-M1 through E1-M16. This change does not alter the concepts discussed previously.}

This equivalence further extends to the case where multiple experts execute concurrently, as shown in Figure \ref{fig:virtualization}(c), where different colors denote micro-slices from distinct experts. Likewise, the expert flow fusion achieves a similar effect to Figure \ref{fig:micro-slice}(b); the specific distribution and the per-chiplet compute duration of a micro-slice (how many tokens are processed) do not affect the aggregate four-expert flow. This “computation and storage-distribution independence” also extends to DDR memory: regardless of storage location, weights can be swept into the dataflow once loaded onto the chip during expert computation. The DDR-load flow can be naturally fused with the D2D flow without detailed pipeline coordination.

Based on the preceding analysis, the expert flow can be realized automatically; the key question is whether it is efficient. Yes—under the following virtualization rules:

\begin{enumerate}[label=\textbf{Rule \arabic*:}, leftmargin=*, nosep]
  \item A micro-slice received in the previous time step is computed immediately in the current time step while simultaneously being transmitted to the next chiplet along the trajectory.

  \item If no micro-slice was received in the previous time step, the chiplet selects any micro-slice from its local storage for immediate computation and transmission.

  \item If there is no next chiplet on the path, the storage occupied by the micro-slice is released immediately after computation.

  \item Chiplets sequentially load the next micro-slice from DDR whenever there is available space. 

  \item DDR controller "sends" micro-slices to the chiplet with the greatest available storage among those processing the expert (Optional).

\end{enumerate}

\textcolor{black}{With these rules, similar ``communication-pattern independence'' can be deduced: under varying ratios of DDR-load to D2D bandwidth, runtime fluctuations, or backpressure, the fused flows self-adapt without detailed pipeline coordination.} Such adaptivity is essential when handling dynamically varying and imbalanced on-chip memory, link utilization, and per-expert token counts, as it enables the expert flow to automatically exploit computation (Rules 1 and 2) and communication (Rules 1 and 4) while minimizing on-chip memory (Rules 2 and 3) resources. Furthermore, because different experts exhibit distinct computation-communication characteristics, these rules allow the fused expert flow to adaptively complement heterogeneous resource usage. \textcolor{black}{In essence, this mechanism functions as a form of hardware-level virtualization, where the shared physical resources—buffers, D2D links, and compute units—are dynamically multiplexed across the logical dataflows of multiple experts. By decoupling the logical trajectory of micro-slices from static physical allocation, the system allows concurrent expert flows to fluidly contend for and utilize the chiplet array's aggregate capacity, resolving local contention and maximizing utilization without the need for cycle-accurate global orchestration.} In our evaluation, we also find that this complementation effectively addresses inter-expert load imbalance and intra-expert differences in the number of tokens that must be processed along the expert trajectory. In other words, FSE-DP \textbf{no longer needs token redistribution}, and all intra-package communication can be simplified to expert point-to-point exchange. \textcolor{black}{Rule 5 is an \emph{optional} optimization and is \emph{not implemented} in our end-to-end system; our ablation study (Section \ref{sec:ablation}) shows that its incremental benefit is limited when other mechanisms (paired-load and flow fusion) are enabled in MoE scenarios, while efficient implementation may require complex metadata tracking in the scheduler or DDR controller.}

\section{Chiplet System with MoE Scheduler}
\label{sec:schedule}

Thanks to virtualization, deployment and scheduling reduce to managing expert trajectories. We propose a scheduling algorithm that orchestrates these trajectories to sustain high hardware utilization under dynamic MoE workloads and asymmetric spatio-temporal chiplet execution patterns. Guided by the virtualization rules, we design and synthesize a lightweight hardware scheduler that realizes this algorithm on a multi-chiplet system equipped with high-speed D2D transceivers.

\subsection{Scheduling Algorithm}

\begin{algorithm}[H]
\caption{Spatiotemporal Trajectory Scheduling Algorithm}
\label{alg:runtime-paired-load-efficient}
\begin{algorithmic}[1]
\REQUIRE
  Expert set $\mathcal{E}$, chiplet set $\mathcal{C}$, idle chiplet set $\mathcal{C}_{\text{idle}} \gets \mathcal{C}$.
\ENSURE
  A dynamic loading sequence of experts with the trajectory $\mathcal{T}_e$ for each expert $e$ that pursues $\mathcal{C}_{\text{idle}} = \varnothing$.
\STATE Sort $\mathcal{E}$ by paired-load policy to an ordered list $\mathcal{E}_{\text{sorted}}$.
\WHILE{not all experts is scheduled ($\mathcal{E}_{\text{sorted}} \neq \varnothing$)}
    \IF{$\mathcal{C}_{\text{idle}} \neq \varnothing$}
        \FOR{each expert pair $(e_1, e_2)$ in $\mathcal{E}_{\text{sorted}}$}
            \STATE Get $\mathcal{T}_e$ from chiplets with tokens for $e$ in $(e_1, e_2)$.
            \IF{there are idle chiplets in the trajectory ($\mathcal{T}_e \cap \mathcal{C}_{\text{idle}} \neq \varnothing$)}
                \STATE Stream $e$'s micro-slice to $c^* \in \mathcal{T}_e \cap \mathcal{C}_{\text{idle}}$.
                \STATE Update idle chiplets: $\mathcal{C}_{\text{idle}} \gets \mathcal{C}_{\text{idle}} \setminus \mathcal{T}_{e}$.
                \STATE Remove $e$ from $\mathcal{E}_{\text{sorted}}$.
                \STATE \textbf{break}
            \ENDIF
            \STATE Rule 4: Pre-load $e$ to any idle buffer. 
        \ENDFOR
    \ENDIF
    
    \STATE When expert $e'$ completes: $\mathcal{C}_{\text{idle}} \gets \mathcal{C}_{\text{idle}} \cup \text{new idle chiplets in } \mathcal{R}_{e'}$.
\ENDWHILE
\end{algorithmic}
\end{algorithm}

Algorithm \ref{alg:runtime-paired-load-efficient} presents a dynamic expert scheduler that assigns experts to chiplets based on token locality and resource availability. Its objective is complete resource utilization—pursuing the condition $\mathcal{C}_{\text{idle}}=\varnothing$—so that all chiplets remain actively engaged in computation. Scheduling begins by ordering experts under a paired-load policy, which prioritizes experts according to complementary computation and communication requirements, yielding an ordered priority queue $\mathcal{E}_{\text{sorted}}$. The main loop repeatedly fetches schedulable experts in priority order and performs resource-aware allocation that respects chiplet availability and trajectory constraints.

For each expert $e$, the algorithm derives a trajectory $\mathcal{T}_e$ representing the path across chiplets that hold cached tokens relevant to $e$. The expert is scheduled only if its trajectory intersects the idle-chiplet set ($\mathcal{T} e \cap \mathcal{C} {\text{idle}} \neq \varnothing$). When such an intersection exists, the algorithm selects an idle chiplet $c^*$ to load a micro-slice of $e$; execution then proceeds automatically under Rules 1–3. Upon successful allocation, the idle set is updated by removing chiplets assigned to the expert’s resource requirements ($\mathcal{R} e$), and the scheduled expert is removed from $\mathcal{E}_{\text{sorted}}$. If the paired expert cannot cover any idle chiplet at the moment, it is pre-loaded on any chiplet with an idle buffer according to Rule 4. Rule 5 is not considered for implementation. When any expert $e'$ completes, chiplets not engaged by other running experts are returned to the idle set, enabling reallocation for subsequent decisions.

\begin{algorithm}[H]
\caption{Token Buffering Algorithm}
\label{alg:token-buffering-disable}
\begin{algorithmic}[1]
\REQUIRE
  \textcolor{black}{Given request $r$, its QoS timer value $T_{\text{QoS}}(r)$, token activation threshold $\theta_{\text{min}}$, count of consecutive forward passes $C_{\text{fw}}(r)$, forward pass threshold to increment the timer $N_{\text{threshold}}$.}

\STATE \textcolor{black}{Let $\mathcal{A}(r)$ be the set of experts activated by $r$ at the current MoE layer, and let $n_e$ be the number of tokens (across all active requests) activating expert $e$.}

\IF{\textcolor{black}{$C_{\text{fw}}(r) \ge N_{\text{threshold}}$}}
    \STATE \textcolor{black}{Increment the QoS timer: $T_{\text{QoS}}(r) \gets T_{\text{QoS}}(r) + 1$.}
    \STATE \textcolor{black}{Reset the forward pass counter: $C_{\text{fw}}(r) \gets 0$.}
\ENDIF

\IF{\textcolor{black}{$(\exists e \in \mathcal{A}(r): n_e < \theta_{\text{min}})$ \AND $T_{\text{QoS}}(r) > 0$}}
    \STATE \textcolor{black}{Defer request $r$ at this MoE layer (token buffering).}
    \STATE \textcolor{black}{Decrement the QoS timer: $T_{\text{QoS}}(r) \gets T_{\text{QoS}}(r) - 1$.}
\ENDIF

\end{algorithmic}
\end{algorithm}

\textcolor{black}{The token-buffering policy (Algorithm \ref{alg:token-buffering-disable}) is applied at each MoE layer boundary \emph{after} gating is computed and \emph{before} scheduling the layer's experts. It decides whether to defer an entire request at that layer, based on (i) whether any of its activated experts are cold under the current per-iteration input-token count and (ii) whether the request has remaining QoS slack. Deferring a request preserves correctness by keeping its intermediate activations and gating results unchanged; the request simply resumes from the same layer in a later iteration.}

\textcolor{black}{For each request, we maintain a QoS timer $T_{\text{QoS}}(r)$. When $T_{\text{QoS}}(r)>0$, token buffering is available for that request. The timer is governed by two rules. First, whenever the consecutive-forward-pass counter $C_{\text{fw}}(r)$ reaches the threshold $N_{\text{threshold}}$, the algorithm increments $T_{\text{QoS}}(r)$ and resets $C_{\text{fw}}(r)$, granting the request one buffering opportunity after a sustained sequence of forward passes. Second, each buffering action decrements $T_{\text{QoS}}(r)$. When buffering triggers, request $r$ is paused at the current MoE layer (its tokens for that layer are not scheduled this iteration), while other requests and experts proceed normally.}

\subsection{Scheduler Hardware}

\begin{figure}[h]
  \centering
  \includegraphics[width=0.86\linewidth]{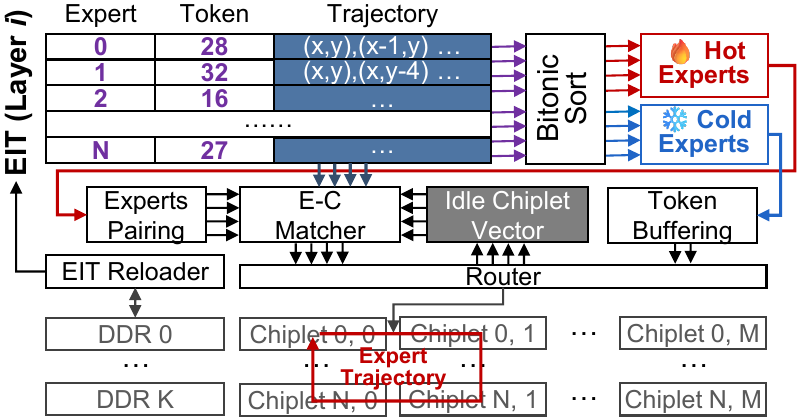}
  \caption{\textbf{Scheduler hardware design.} The scheduler is implemented in the IO die for task allocation.}
  \label{fig:hw}
\end{figure}

To enable efficient execution of the proposed scheduling algorithms in real-world multi-chiplet systems, we design a dedicated hardware scheduler that implements trajectory-aware dynamic scheduling and token buffering in silicon, as shown in Figure \ref{fig:hw}. The implementation comprises key components that operate synergistically to deliver low-latency scheduling decisions while sustaining high resource utilization.

\textbf{Expert Information Table (EIT)} serves as a low-latency lookup that maps expert identifiers to trajectory masks. Implemented in single-cycle SRAM with expert IDs as keys, it stores the relative trajectory and the number of activating tokens as values. This structure enables immediate trajectory resolution without iterative searches, which reduces scheduling overhead. To classify hot and cold experts, a bitonic sorter performs parallel sorting of all experts by their token counts. Hot experts are paired as illustrated in Figure \ref{fig:expert_laod}, whereas tokens associated with cold experts are diverted to token buffering.

\textbf{Idle Chiplet Vector (ICV)} tracks chiplet availability in real time via a dedicated N-bit register bank. It supports concurrent reads for scheduling decisions and asynchronous writes triggered by expert-completion events. Efficient updates are realized with bit-wise operations: allocation uses AND–NOT masking with trajectory patterns, while completion-driven releases use OR with completion masks.

\textbf{Expert–Chiplet Matcher (E–C Matcher)} assigns computing chiplets and their communication paths based on the ICV and the trajectory required for each expert’s execution.

We synthesized this scheduler architecture and integrated \textcolor{black}{its RTL} into a fabricated 4‑chiplet prototype. The complete scheduler occupies only 0.43 $mm^2$, as reported by Synopsys Design Compiler in UMC 28‑nm technology, and achieves sub‑microsecond scheduling latency under typical expert configurations. This implementation demonstrates the practical feasibility of our algorithms and enables comprehensive evaluation of the proposed FSE‑DP framework. Detailed performance results are presented in Section \ref{sec:evaluation}.

\subsection{Hardware Implementation of Accelerator}

Finally, we introduce the hardware architecture and the accelerator design within each chiplet. Figure \ref{fig:introduction} shows that each chiplet comprises a PE array and SRAM-based on-chip memory \cite{10454441}. For each computing die, the basic architecture is a transformer-oriented accelerator architecture, similar to \cite{10904499}. Each compute die integrates a PE array for linear operations, a non-linear unit (NLU), UCIe IP modules, and a data movement unit (DMU) for data transfer and format conversion. In addition, each die incorporates a router that receives task sequences issued by the scheduler and enables communication with other compute dies. After the scheduler assigns tasks, the compute die loads weights from DDR or from peer dies into private memory for expert inference. For programming, each chiplet internally stores a static instruction sequence for transformer operations, but the communication routing table is generated in real time. In other words, whether data are sent/received to/from DDR or another die is determined by the trajectory provided by the real-time scheduler.

\section{Evaluation}
\label{sec:evaluation}

\subsection{Experimental Setup}

\textbf{Models and Datasets.} We evaluate FSE-DP on four representative MoE models: Phi-3.5-MoE~\cite{team2024phi}, Yuan2.0-M32~\cite{wu2024yuan}, DeepSeek-MoE~\cite{rajbhandari2022deepspeed}, and Qwen3-30B-A3B~\cite{yang2025qwen3}. These models respectively contain 16, 32, 64, and 128 experts per layer, offering a diverse spectrum of model scales; detailed specifications are listed in Table~\ref{tab:model-params}. For evaluation, we employ two widely used language-modeling benchmarks, Wikitext-2~\cite{merity2016pointer} and C4~\cite{raffel2020exploring}. \textcolor{black}{Our goal is to stress MoE activation skew and off-chip expert fetch behavior under controlled per-iteration input-token counts; in addition, we use WinoGrande in our motivation profiling (Section \ref{sec:background}) to show that the long-tail activation pattern persists across task types.}

\begin{table}[htbp]
\centering
\setlength{\abovecaptionskip}{2pt}
\caption{\textcolor{black}{Hardware and Model Configurations for Evaluation}}
\label{tab:model-params}

\begin{tabular}{cl}
\hline
\textbf{Component} & \textbf{Specification} \\
\hline
& DDR3-1600 4$\times$25.6 GB/s, 800 MHz \\
NoP \& DDR& 2D Mesh, Multiple UCIe D2D IPs: 288 GB/s  \\
& 4-24 Gbps/pin, FDI-to-FDI Latency: 4.02 ns\\
\hline
  & Samsung 5nm 1P13M CMOS, 800 MHz \\
Compute Die & 2048 MACs, 0.675-0.9 V, 736.5 -2187 mW \\
  & 2.69mm$\times$4.72mm, 4.865 TOPS\\
\hline
\end{tabular}
\label{tab:hardware_config}

\vspace{0.2cm}

\begin{tabular}{cllllll}
\hline
\textbf{Model} & \textbf{$D_{model}$} & \textbf{$D_{ffn}$} & \textbf{${E}$} & \textbf{${E}^{act}$} & \textbf{Head} & \textbf{Para.} \\ \hline
Phi-3.5        & 4096                 & 3200               & 16             & 2                    & 32             & 41.9B          \\
Yuan2.0-M32    & 2048                 & 4096               & 32             & 2                    & 16             & 40B            \\
DeepSeek-MoE   & 2048                 & 1408               & 64             & 6+2                  & 16             & 16.4B          \\
Qwen3-A3B  & 2048                 & 768                & 128            & 8                    & 32             & 30B            \\ \hline
\end{tabular}
\label{tab:model_config}

\end{table}

\textbf{Baseline.} We evaluate two representative baselines for MoE inference. EP is the de facto method in this field, distributing experts across devices (each device hosting a distinct subset of experts) and routing tokens via all-to-all; its convenient implementation and relatively acceptable performance serve as a widely adopted reference. Hydra is a software–hardware co-designed scheme optimized for multi-chiplet systems; here, we isolate its optimization of EP: it exploits cross-layer expert popularity to relocate experts and reduce inter-chiplet communication~\cite{he2025hydra}. Using both allows us to compare against the standard EP paradigm and a state-of-the-art chiplet-specialized distributed strategy.

\begin{figure*}[h]
  \centering
  \includegraphics[width=1\linewidth]{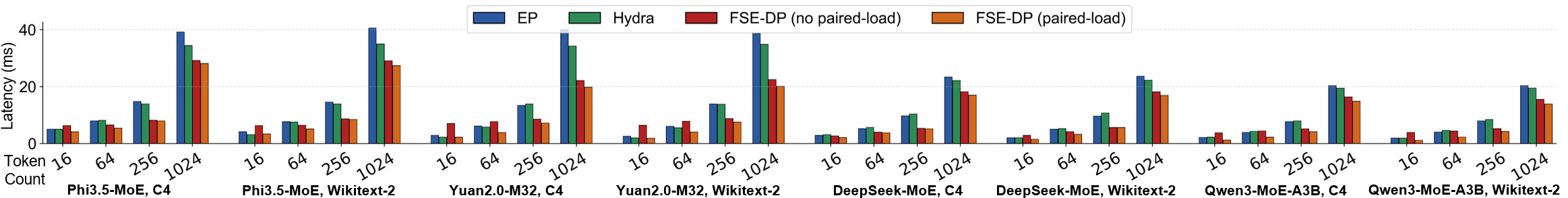}
  \caption{\textbf{Single MoE layer latency with different models, datasets and input token counts.} }
  \label{fig:utilization_bar_8groups}
\end{figure*}

\begin{figure}[h]
  \centering
  \includegraphics[width=0.5\linewidth]{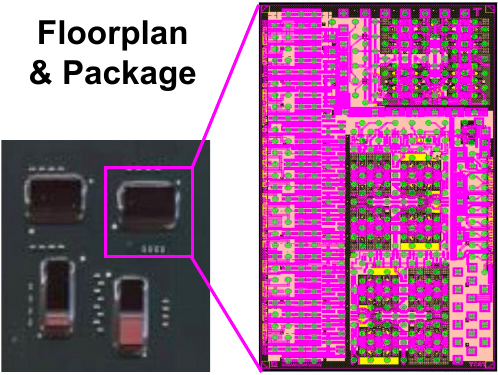}
  \caption{\textbf{Chiplet floorplan and package photo} }
  \label{fig:chip}
\end{figure}

\textbf{Implementation and configuration.} \textcolor{black}{In this work, evaluation results are produced by cycle-accurate simulators with RTL-synthesized expert trajectory scheduler of a tape-out 2×2 5nm test chip that executes expert activation and execution for the aforementioned networks on datasets.} We further use the test chip and RTL-synthesized performance to calibrate behavioral-level simulations\cite{qu2025mldse} for other hardware configurations during design space exploration (DSE). Figure \ref{fig:chip} illustrates our prototype multi-chiplet system, and Table \ref{tab:hardware_config} presents the basic specifications based on this chip.

\textcolor{black}{Given a low-batch scenario, the number of concurrent requests is small, and contexts from different requests (prefill and decode phases) are mixed for inference. Accordingly, we quantify the ``effective batch'' using \emph{tokens-per-iteration} (micro-batch tokens): the number of \emph{input} tokens aggregated across a small set of concurrent requests and processed in one forward scheduling iteration. We report fixed tokens-per-iteration settings---16, 64, 256, and 1024---with requests sampled from the Wikitext-2 and C4 datasets. This avoids ambiguity in request-count ``batch size'' under mixed prefill/decode, variable context lengths, and chunked prefill, and directly reflects system pressure and weight-reuse intensity. These values are not the output length nor the context length of a single request.} When involving token buffering, we configure three slackness levels of 10\%, 20\%, and 30\%, which denote the fraction of an iteration that a request is allowed to be deferred at a MoE layer boundary, emulating diverse QoS.


\textbf{Methodology.} We structure our evaluation into four phases. First, since most MoE in LLMs is applied within FFN blocks while attention remains dense, we isolate expert computation to avoid confounding factors from attention implementations and benchmark EP, Hydra, and FSE-DP with and without paired load on a single FFN MoE layer. Next, we implement a basic attention scheduling scheme and obtain end-to-end performance across 100 consecutive forward iterations; we also conduct five ablation configurations to quantify the contribution of individual optimizations under varying conditions. Token buffering, which involves cross-iteration operations, is enabled only in the end-to-end set. Then, we perform DSE to study FSE-DP sensitivity to on-chip memory, off-chip and D2D bandwidth. Finally, we assess the scalability of our spatiotemporal scheduling on different arrays ($3\times 3$, $4\times 4$). \textcolor{black}{In the evaluation, we schedule an expert trajectory as a ring, using next-hop forwarding for implementation simplicity. The ring is a \emph{logical} route and is not tied to a physical ring topology. When the array is larger than 2×2, we use a 2D-mesh interconnect to apply multiple ring trajectories concurrently.}

\subsection{Isolated Expert-Compute Performance}

Firstly, we focus on the performance of the MoE part only. Figure~\ref{fig:utilization_bar_8groups} illustrates the latency results, averaged across all layers of the network. In most configurations, FSE-DP achieves the lowest latency. Specifically, when the token count is relatively low, the paired-load mechanism yields significant improvements. As the token count increases, each expert performs more computation and the DDR bottleneck gradually eases; the performance advantage of FSE-DP then stems from its full utilization of the chiplet's D2D bandwidth compared with other solutions. Hydra mainly focuses on optimizing the collective communication of tokens, which is less beneficial in low-batch, high-D2D-bandwidth scenarios; consequently, it shows no obvious improvement over EP.

\begin{figure}[h]
  \centering
  \includegraphics[width=1\linewidth]{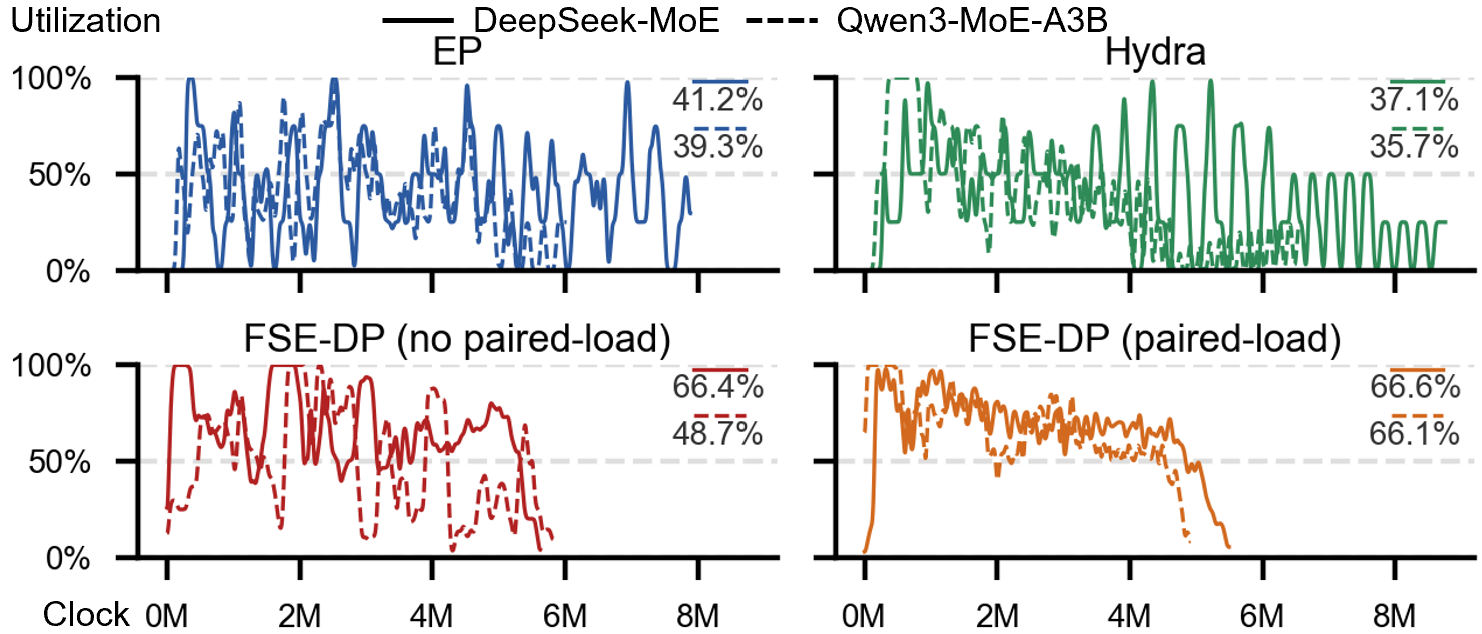}
  \caption{\textbf{Utilization fluctuation during inference of one layer.} }
  \label{fig:utilization_lines_4groups}
\end{figure}

Figure~\ref{fig:utilization_lines_4groups} analyzes the performance differences among the four scheduling approaches from a temporal perspective. The utilization curves reveal the source of the performance gain: FSE-DP exhibits much smaller performance fluctuations than EP and Hydra. These benefits arise from expert sharing and dynamic trajectories that avoid spatiotemporal congestion on both bandwidth and compute resources during the inference of complementary experts.

\begin{figure}[h]
  \centering
  \includegraphics[width=1\linewidth]{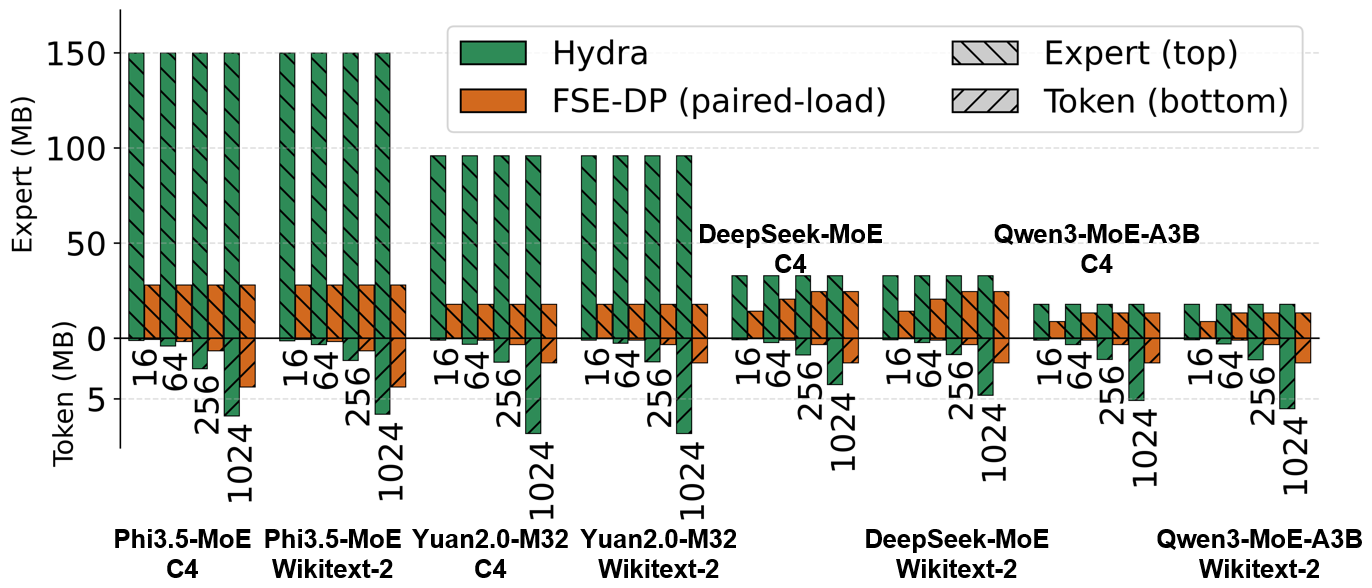}
  \caption{\textbf{On-chip memory usage of different models.} }
  \label{fig:max_buffer_bar_8groups}
\end{figure}

Figure~\ref{fig:max_buffer_bar_8groups} illustrates the on-chip memory usage of the different models that achieve the performance reported in the previous Figure~\ref{fig:utilization_bar_8groups}. Compared with EP and Hydra, FSE-DP shows a significant reduction in memory cost, especially when the expert dimension is large. FSE-DP achieves this by sharding each expert into micro-slices and applying Rules 1–4 to ensure that a micro-slice is rapidly released from the package. Consequently, we compress the on-chip memory overhead of the multi-chiplet system to less than 32 MB—about one-fifth of that required by EP and Hydra. Without token replication, token storage usage is also reduced. Note that the fine-grained expert flow endows FSE-DP with a degree of elasticity in the on-chip buffer: smaller buffer sizes are permissible at the cost of some performance loss, while larger buffers can further improve performance (as discussed in DSE section).

\begin{figure}[h]
  \centering
  \includegraphics[width=1\linewidth]{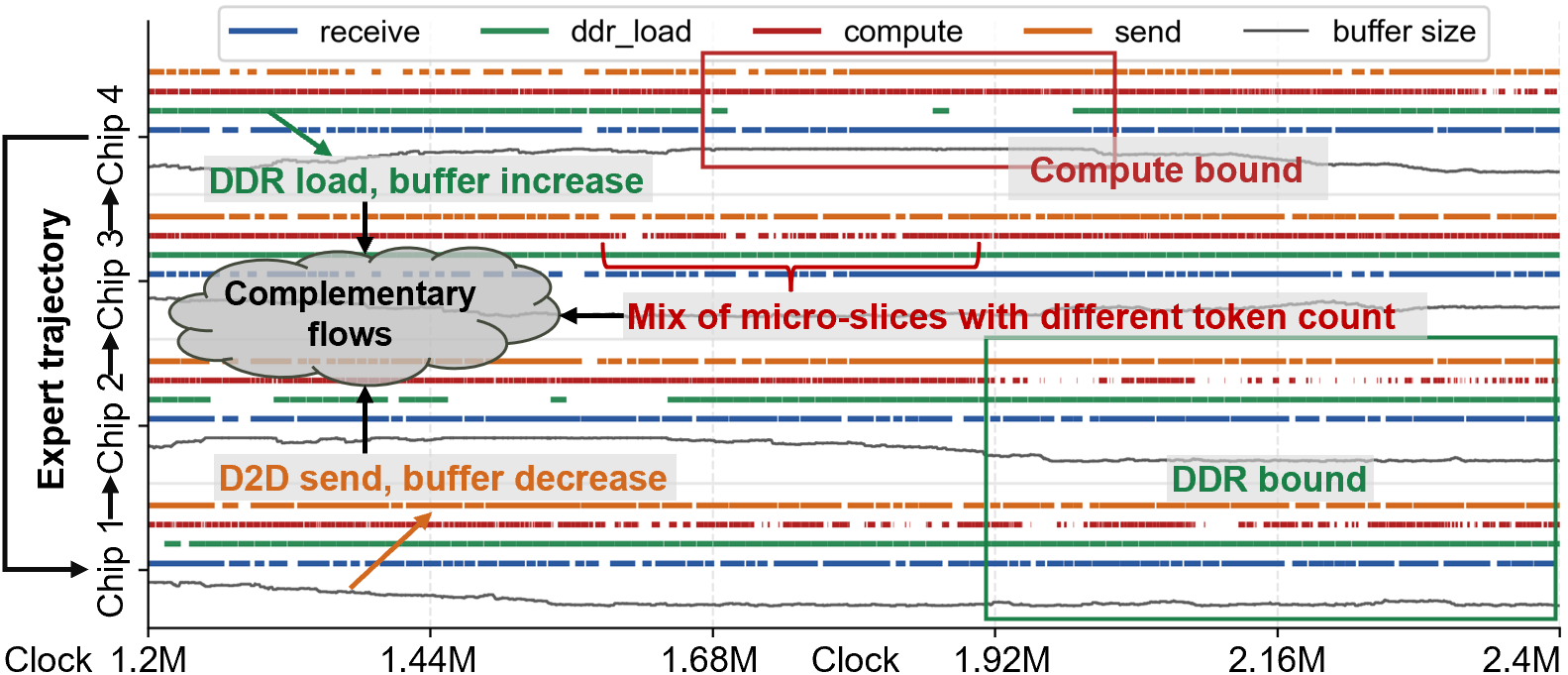}
  \caption{\textbf{Activity timeline of expert trajectories across chiplets under FSE-DP (paired load). Qwen3-MoE, C4 with 256 input tokens, a runtime snapshot segment.}}
  \label{fig:activity_timeline}
\end{figure}

To further illustrate the efficacy of complementary flows, we decompose the activities across four chiplets in Figure~\ref{fig:activity_timeline}. Given the substantial overlap of D2D communication (send/receive), DDR load, and computation, we depict them as a clock-aligned timeline. Two complementary mechanisms boost performance. First, because the workload assigned to each chiplet for every expert micro-slice varies dynamically, the on-chip micro-slice buffer acts as an elastic reservoir that absorbs mismatches between D2D traffic and DDR access, keeping both interfaces highly utilized. Second, the interleaving of heterogeneous expert flows injects computations of varying durations into each chiplet, balancing computation-bound and communication-bound phases. Still, when an expert’s demand exceeds the adaptive ceiling, a resource bound still occurs.

\subsection{End-to-End Evaluation with Ablation Studies}
\label{sec:ablation}

\begin{figure}[h]
  \centering
  \includegraphics[width=1\linewidth]{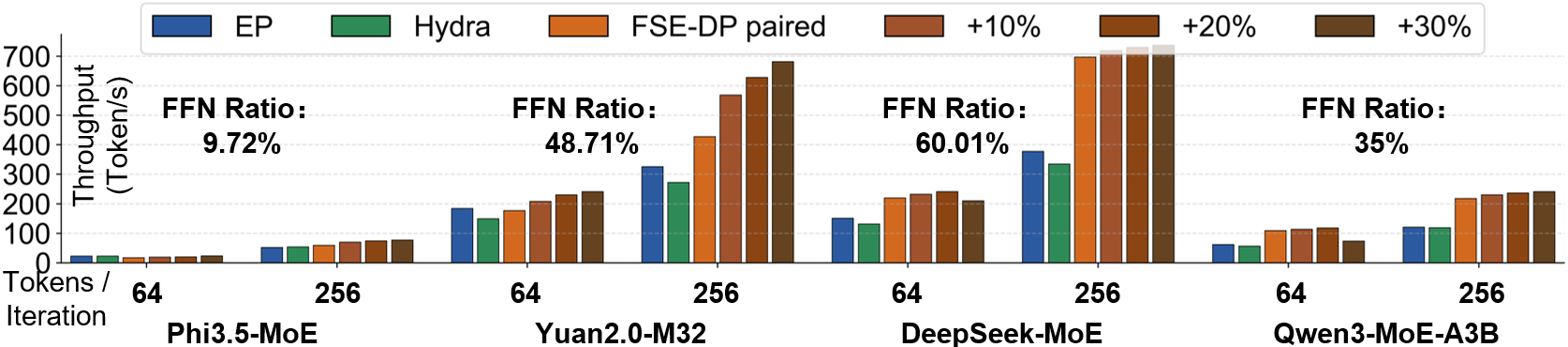}
  \caption{\textbf{End-to-end throughput comparison across model–dataset combinations.} +10\% means paired-load policy with 10\% token buffering slackness.}
  \label{fig:buffering_groups_bars}
\end{figure}

Next, we evaluate end-to-end performance by combining the attention phase with 100 forward iterations of the aforementioned workloads and perform ablation studies. For attention, we perform head parallelism on different chiplets. Figure~\ref{fig:buffering_groups_bars} compares different strategies, including token buffering. FSE-DP with moderate buffering slack significantly improves throughput; however, excessive slack can degrade performance. Token buffering deliberately delays the processing of a request, and when the total token count is small, the resulting reduction in compute volume amplifies the data-transfer bottleneck, yielding no net gain in utilization. Note also that in networks such as Phi3.5-MoE the FFN fraction is small, so MoE-centric optimizations have limited impact.

\begin{figure}[h]
  \centering
  \includegraphics[width=1\linewidth]{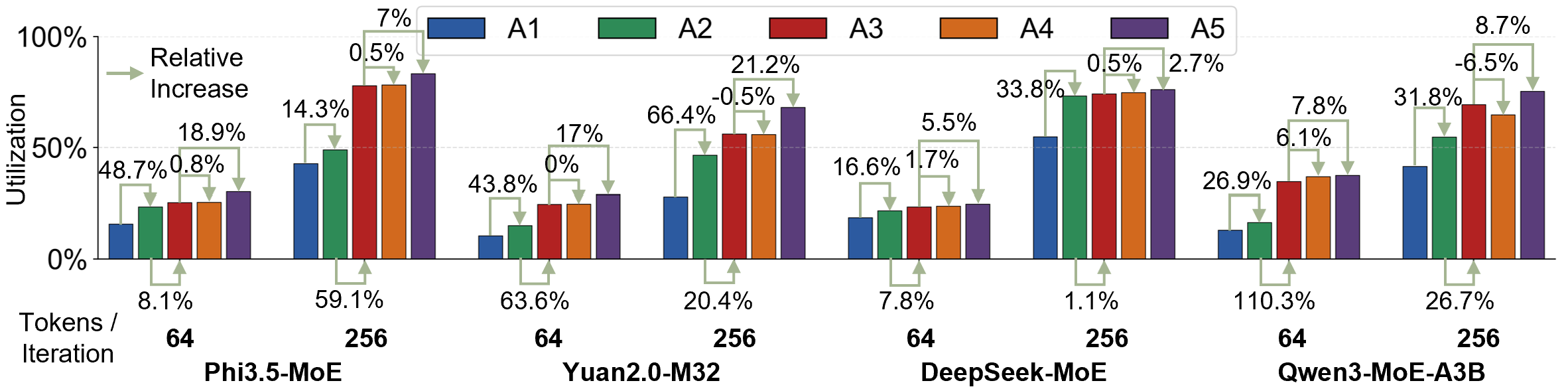}
  \caption{\textbf{Ablation study on key design knobs of FSE-DP.}}
  \label{fig:ablation_groups_bars}
\end{figure}

We define five ablation configurations. A1: naive FSE-DP without fine-grained flows. A2: FSE-DP with fine-grained flows governed by Rules 1--4. A3: A2 + paired-load policy. A4: A3 + Rule 5 (\emph{optional}, excluded from our main end-to-end implementation). A5: A3 + 20\% token buffering. Note that A2 and A3 are the configurations adopted in the preceding experiments. Figure~\ref{fig:ablation_groups_bars} reports the utilization achieved by each setup. Both paired-load and token buffering significantly improve performance, whereas Rule 5 yields only marginal gains.

\subsection{Design Space Exploration with Sensitivity Analysis}
\label{sec:dse}

\textcolor{black}{We investigate the sensitivity of hardware configuration to FSE-DP performance. We impose two constraints in the search process:}

\begin{equation}
    \lceil \frac{BW_{D2D}}{BW_{UCIe}} \rceil A_{UCIe} + A_{Compute} + A_{Buffer} \leq A_{th}
\end{equation}

\begin{equation}
    P_{Compute} + P_{D2D} + P_{DDR} \leq P_{th}
\end{equation}

\begin{figure}[h]
  \centering
  \includegraphics[width=1\linewidth]{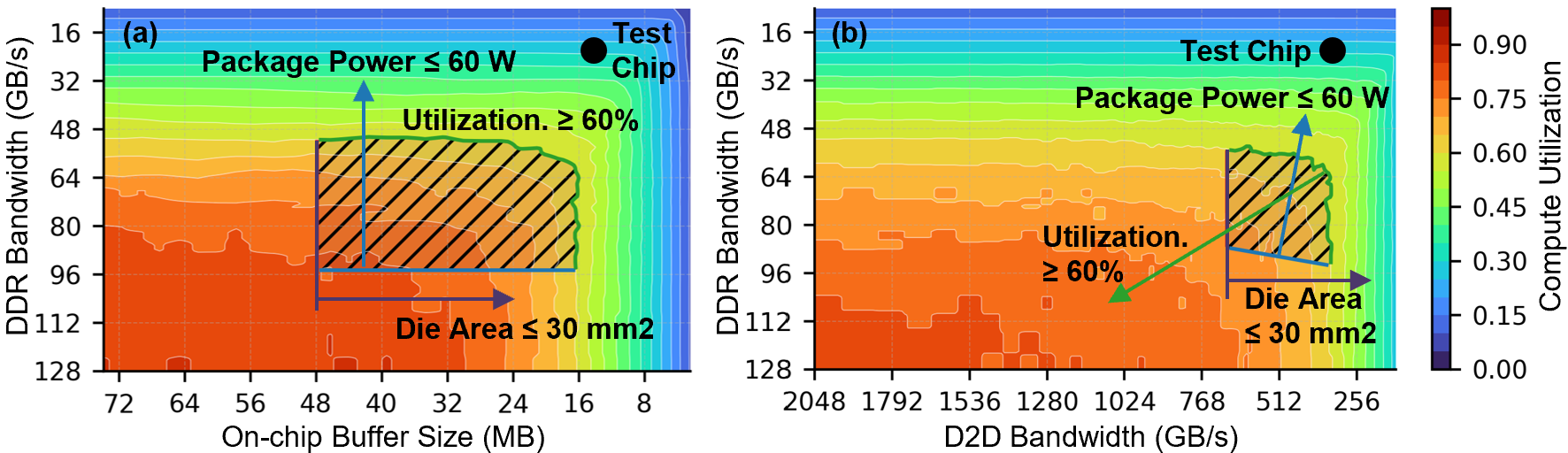}
  \caption{\textbf{DSE.} Qwen3-MoE-A3B, C4, 64 input tokens. The star is the position of our test chip. Much larger scales are preferred in practice. (a) Fix D2D bandwidth at 288 GB/s. (b) Fix buffer size at 14MB.}
  \label{fig:heatmap_dse}
\end{figure}

They respectively represent the area constraints of individual chiplets and the peak power consumption constraints of the entire package. Figure \ref{fig:heatmap_dse} illustrates the evaluation results and marks the domain satisfying the constraints as shaded.

In Figure \ref{fig:heatmap_dse}(a), we fix the D2D bandwidth and then analyze the relationship between the on-chip buffer size and the DDR bandwidth. When we set the upper limit of the die area to 30 $mm^2$, the total power consumption of the entire package is less than 60W. The results show that in order to achieve a utilization rate higher than 60\%, 48 GB/s of DDR bandwidth per die and 16MB or more of on-chip memory are required.

Figure \ref{fig:heatmap_dse}(b) further analyzes the trade-off between DDR bandwidth and D2D bandwidth. We fix the on-chip memory to only 14MB, which is outside the shaded area in Figure \ref{fig:heatmap_dse}(a). The results show that in this case, the area that can meet the constraints and performance requirements is very limited. Moreover, a very high D2D bandwidth is required to compensate for the capacity of the on-chip memory, even up to 512GB/s, which is equivalent to 3 UCIe ($\times$ 32) modules. This still poses very significant design challenges.

In conclusion, the lesson this experiment teaches us is as follows: For ideas similar to T10\cite{t10}, where \textbf{trading communication performance for DDR bandwidth, a relatively large on-chip memory capacity is necessary as a guarantee in multi-chiplet MoE inference}.

\begin{figure}[h]
  \centering
  \includegraphics[width=1\linewidth]{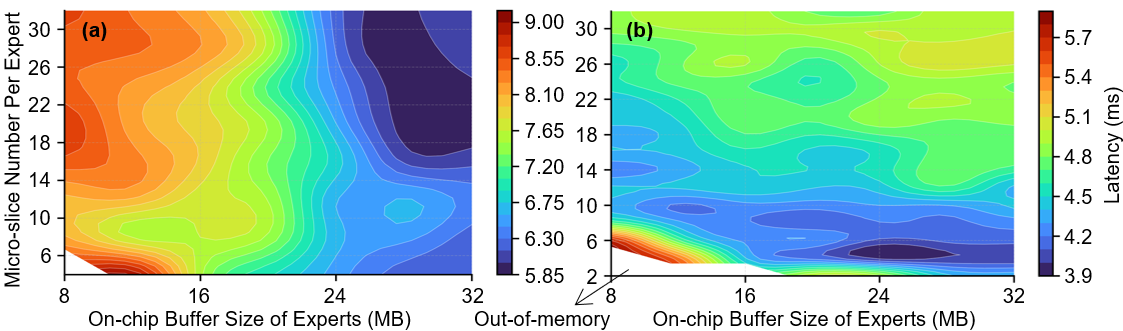}
  \caption{\textcolor{black}{\textbf{Granularity sensitivity.} Latency heatmap over on-chip expert weight storage size and micro-slice number, evaluated on (a) Phi-3.5 and (b) Qwen3-MoE-A3B using C4.}}
  \label{fig:granularity_heatmap}
\end{figure}

\textcolor{black}{Complementary to the DSE results on memory and bandwidth, we further evaluate how \emph{micro-slice granularity} and the available on-chip expert storage affect end-to-end latency. Figure \ref{fig:granularity_heatmap} reports a latency heatmap for Phi-3.5 and Qwen3-MoE-A3B on C4. Because Phi-3.5 has larger experts while Qwen3-MoE-A3B uses smaller per-expert models, the two exhibit different sensitivities to the micro-slice granularity. When micro-slices are overly fine-grained, the per-micro-slice control overhead cannot be overlapped by the computation within each micro-slice, making control cost a considerable performance factor. This effect is more pronounced for models with smaller experts (e.g., Qwen3-MoE-A3B). Empirically, a micro-slice number below 10 is preferred. In contrast, for Phi-3.5, performance is more strongly influenced by the on-chip buffer size, where increasing on-chip memory yields a clearer speedup. Conversely, overly coarse granularity prevents our method from leveraging fine-grained, adaptive dataflow and can also degrade performance. As a result, increasing the micro-slice number may first improve and then worsen performance. Due to the coupling among multiple factors and the inherent stochasticity of MoE routing, these trends may not always appear clearly in end-to-end measurements.}

\subsection{Scalability}

\begin{figure}[h]
  \centering
  \includegraphics[width=1\linewidth]{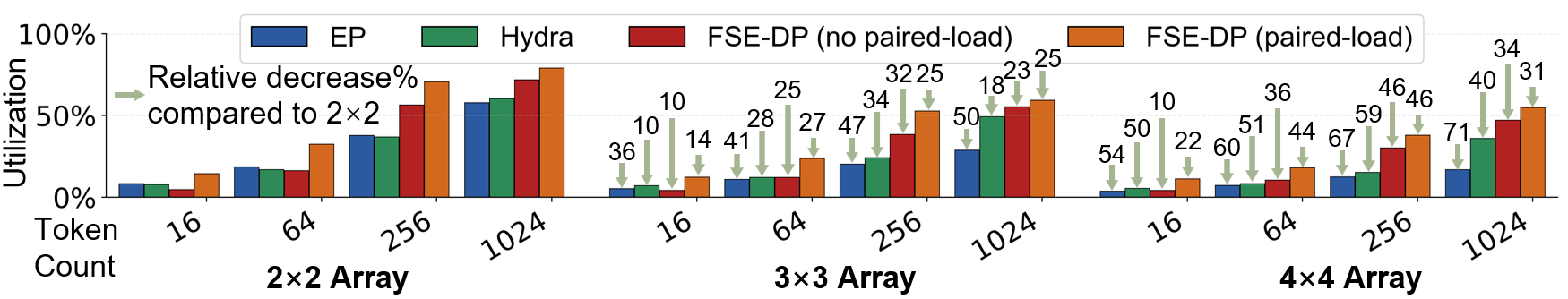}
  \caption{\textcolor{black}{\textbf{Scalability (utilization) evaluation} based on Qwen3-MoE-A3B, C4.}}
  \label{fig:utilization_chiparrays_qwen_c4}
\end{figure}

\textcolor{black}{We then analyzed scalability from $2\times2$ to $4\times4$ chiplet arrays in Figure \ref{fig:utilization_chiparrays_qwen_c4}. This figure reports \emph{utilization}. Higher utilization generally correlates with lower expert-layer latency under fixed frequency, but utilization alone could mask latency increases caused by additional hops and congestion at larger scales. Our simulator models D2D transfer time along the chosen trajectories, so the utilization trend reflects additional idle cycles due to inter-chiplet movement.} As the array grows, EP's efficiency decreases significantly. In contrast, Hydra improves scalability by optimizing collective communication. FSE-DP also scales well: compared with EP and Hydra, the utilization of FSE-DP with only point-to-point communication decreases significantly less in larger arrays, benefiting from trajectory-aware scheduling and the avoidance of all-to-all.

\section{Conclusion}

\textcolor{black}{This paper presents FSE-DP, a multi-chiplet parallel strategy for low-batch MoE inference, which effectively addresses key challenges in edge deployment such as on-chip memory constraints, off-chip bandwidth bottlenecks, and load imbalance through dynamic expert trajectory scheduling. The fine-grained dataflow opens the opportunity for adaptive complementary resource utilization. Our method can directly apply to MoT-style designs where attention blocks are expertized; beyond expertized networks, the virtualization method can be extended to a broader programming model for other dynamic workloads, such as KV-Cache management, in the future. However, our method imposes requirements on hardware: sufficient D2D link efficiency and fine-grained capabilities in computation, memory access, and communication, which limits the range of platforms on which it can be applied. We design and synthesize a lightweight hardware scheduler and evaluate FSE-DP using an RTL cycle-accurate simulator of a $2\times2$ 5‑nm test chip. Experimental results show that it outperforms existing schemes across models and tokens-per-iteration configurations, improving latency and on-chip memory overhead.}



\bibliographystyle{IEEEtranS}
\bibliography{sample-base}

@article{zhang2024m2m,
  title={M2M: A Fine-Grained Mapping Framework to Accelerate Multiple DNNs on a Multi-Chiplet Architecture},
  author={Zhang, Jinming and Wang, Xuyan and Ye, Yaoyao and Lyu, Dongxu and Xiong, Guojie and Xu, Ningyi and Lian, Yong and He, Guanghui},
  journal={IEEE Transactions on Very Large Scale Integration (VLSI) Systems},
  year={2024},
  publisher={IEEE}
}

@inproceedings{shao2019simba,
  title={Simba: Scaling deep-learning inference with multi-chip-module-based architecture},
  author={Shao, Yakun Sophia and Clemons, Jason and Venkatesan, Rangharajan and Zimmer, Brian and Fojtik, Matthew and Jiang, Nan and Keller, Ben and Klinefelter, Alicia and Pinckney, Nathaniel and Raina, Priyanka and others},
  booktitle={Proceedings of the 52nd annual IEEE/ACM international symposium on microarchitecture},
  pages={14--27},
  year={2019}
}

@inproceedings{tan2021nn,
  title={Nn-baton: Dnn workload orchestration and chiplet granularity exploration for multichip accelerators},
  author={Tan, Zhanhong and Cai, Hongyu and Dong, Runpei and Ma, Kaisheng},
  booktitle={2021 ACM/IEEE 48th Annual International Symposium on Computer Architecture (ISCA)},
  pages={1013--1026},
  year={2021},
  organization={IEEE}
}

@inproceedings{lin2024hex,
  title={HEX-SIM: Evaluating Multi-modal Large Language Models on Multi-chiplet NPUs},
  author={Lin, Xinquan and Xu, Haobo and Han, Yinhe and Gan, Yiming},
  booktitle={2024 IEEE International Symposium on Workload Characterization (IISWC)},
  pages={108--120},
  year={2024},
  organization={IEEE}
}

@article{fedus2022switch,
  title={Switch transformers: Scaling to trillion parameter models with simple and efficient sparsity},
  author={Fedus, William and Zoph, Barret and Shazeer, Noam},
  journal={Journal of Machine Learning Research},
  volume={23},
  number={120},
  pages={1--39},
  year={2022}
}

@article{jiang2024mixtral,
  title={Mixtral of experts},
  author={Jiang, Albert Q and Sablayrolles, Alexandre and Roux, Antoine and Mensch, Arthur and Savary, Blanche and Bamford, Chris and Chaplot, Devendra Singh and Casas, Diego de las and Hanna, Emma Bou and Bressand, Florian and others},
  journal={arXiv preprint arXiv:2401.04088},
  year={2024}
}

@article{yang2025qwen3,
  title={Qwen3 technical report},
  author={Yang, An and Li, Anfeng and Yang, Baosong and Zhang, Beichen and Hui, Binyuan and Zheng, Bo and Yu, Bowen and Gao, Chang and Huang, Chengen and Lv, Chenxu and others},
  journal={arXiv preprint arXiv:2505.09388},
  year={2025}
}

@article{wei2024deepseek-ocr,
  title={DeepSeek-OCR: Contexts Optical Compression},
  author={Wei, Haoran and Sun, Yaofeng and Li, Yukun},
  journal={arXiv preprint arXiv:2510.18234},
  year={2025}
}

@article{wang2024auxiliary,
  title={Auxiliary-loss-free load balancing strategy for mixture-of-experts},
  author={Wang, Lean and Gao, Huazuo and Zhao, Chenggang and Sun, Xu and Dai, Damai},
  journal={arXiv preprint arXiv:2408.15664},
  year={2024}
}

@article{liu2025expert,
  title={Expert-as-a-Service: Towards Efficient, Scalable, and Robust Large-scale MoE Serving},
  author={Liu, Ziming and Tian, Boyu and Wang, Guoteng and Jiang, Zhen and Sun, Peng and Han, Zhenhua and Tang, Tian and Hu, Xiaohe and Jia, Yanmin and Zhang, Yan and others},
  journal={arXiv preprint arXiv:2509.17863},
  year={2025}
}

@inproceedings{srinivasa2025300mb,
  title={A 300MB SRAM, 20Tb/s Bandwidth Scalable Heterogenous 2.5 D System Inferencing Simultaneous Streams Across 20 Chiplets with Workload-Dependent Configurations},
  author={Srinivasa, Srivatsa Rangachar and Kurian, Dileep and Aseron, Paolo and Budhkar, Prerna and Radhakrishnan, Arun and Lopez, Alejandro Cardenas and Sundaram, Jainaveen and Honkote, Vinayak and Azarenkov, Leonid and Lake, Daniel and others},
  booktitle={2025 IEEE International Solid-State Circuits Conference (ISSCC)},
  volume={68},
  pages={50--52},
  year={2025},
  organization={IEEE}
}

@inproceedings{singh2023hybrid,
  title={A hybrid tensor-expert-data parallelism approach to optimize mixture-of-experts training},
  author={Singh, Siddharth and Ruwase, Olatunji and Awan, Ammar Ahmad and Rajbhandari, Samyam and He, Yuxiong and Bhatele, Abhinav},
  booktitle={Proceedings of the 37th International Conference on Supercomputing},
  pages={203--214},
  year={2023}
}

@article{li2025moe,
  title={The MoE-Empowered Edge LLMs Deployment: Architecture, Challenges, and Opportunities},
  author={Li, Ning and Guo, Song and Zhang, Tuo and Li, Muqing and Hong, Zicong and Zhou, Qihua and Yuan, Xin and Zhang, Haijun},
  journal={arXiv preprint arXiv:2502.08381},
  year={2025}
}

@article{chen2023pipeline,
  title={Pipeline moe: A flexible moe implementation with pipeline parallelism},
  author={Chen, Xin and Zhang, Hengheng and Gu, Xiaotao and Bi, Kaifeng and Xie, Lingxi and Tian, Qi},
  journal={arXiv preprint arXiv:2304.11414},
  year={2023}
}

@article{hwang2023tutel,
  title={Tutel: Adaptive mixture-of-experts at scale},
  author={Hwang, Changho and Cui, Wei and Xiong, Yifan and Yang, Ziyue and Liu, Ze and Hu, Han and Wang, Zilong and Salas, Rafael and Jose, Jithin and Ram, Prabhat and others},
  journal={Proceedings of Machine Learning and Systems},
  volume={5},
  pages={269--287},
  year={2023}
}

@article{yang2025hybrid,
      title={HybridEP: Scaling Expert Parallelism to Cross-Datacenter Scenario via Hybrid Expert/Data Transmission}, 
      author={Weihao, Yang and Hao, Huang and Donglei, Wu and Ningke, Li and Yanqi, Pan and Qiyang, Zheng and Wen, Xia and Shiyi, Li and Qiang, Wang},
      journal={arXiv preprint arXiv:2510.19470},
      year={2025}
}

@inproceedings{t10,
author = {Liu, Yiqi and Xue, Yuqi and Cheng, Yu and Ma, Lingxiao and Miao, Ziming and Xue, Jilong and Huang, Jian},
title = {Scaling Deep Learning Computation over the Inter-Core Connected Intelligence Processor with T10},
year = {2024},
isbn = {9798400712517},
publisher = {Association for Computing Machinery},
address = {New York, NY, USA},
url = {https://doi.org/10.1145/3694715.3695955},
doi = {10.1145/3694715.3695955},
booktitle = {Proceedings of the ACM SIGOPS 30th Symposium on Operating Systems Principles},
pages = {505–521},
numpages = {17},
location = {Austin, TX, USA},
series = {SOSP '24}
}

@INPROCEEDINGS{10454441,
  author={Smith, Alan and Chapman, Eric and Patel, Chintan and Swaminathan, Raja and Wuu, John and Huang, Tyrone and Jung, Wonjun and Kaganov, Alexander and McIntyre, Hugh and Mangaser, Ramon},
  booktitle={2024 IEEE International Solid-State Circuits Conference (ISSCC)}, 
  title={11.1 AMD InstinctTM MI300 Series Modular Chiplet Package – HPC and AI Accelerator for Exa-Class Systems}, 
  year={2024},
  volume={67},
  number={},
  pages={490-492},
  doi={10.1109/ISSCC49657.2024.10454441}}

@INPROCEEDINGS{10904499,
  author={Dong, Pingcheng and Tan, Yonghao and Liu, Xuejiao and Luo, Peng and Liu, Yu and Liang, Luhong and Zhou, Yitong and Pang, Di and Yung, Man-To and Zhang, Dong and Huang, Xijie and Liu, Shih-Yang and Wu, Yongkun and Tian, Fengshi and Tsui, Chi-Ying and Tu, Fengbin and Cheng, Kwang-Ting},
  booktitle={2025 IEEE International Solid-State Circuits Conference (ISSCC)}, 
  title={A 28nm 0.22$\mu$J/Token Memory-Compute-Intensity-Aware CNN-Transformer Accelerator with Hybrid-Attention-Based Layer-Fusion and Cascaded Pruning for Semantic-Segmentation}, 
  year={2025},
  volume={68},
  number={},
  pages={01-03},
  doi={10.1109/ISSCC49661.2025.10904499}}

@article{gupta2024lynx,
  title={Lynx: Enabling efficient moe inference through dynamic batch-aware expert selection},
  author={Gupta, Vima and Sinha, Kartik and Gavrilovska, Ada and Iyer, Anand Padmanabha},
  journal={arXiv preprint arXiv:2411.08982},
  year={2024}
}

@inproceedings{gao2019tangram,
  title={Tangram: Optimized coarse-grained dataflow for scalable nn accelerators},
  author={Gao, Mingyu and Yang, Xuan and Pu, Jing and Horowitz, Mark and Kozyrakis, Christos},
  booktitle={Proceedings of the Twenty-Fourth International Conference on Architectural Support for Programming Languages and Operating Systems},
  pages={807--820},
  year={2019}
}

@inproceedings{he2025waferllm,
  title={$\{$WaferLLM$\}$: Large Language Model Inference at Wafer Scale},
  author={He, Congjie and Huang, Yeqi and Mu, Pei and Miao, Ziming and Xue, Jilong and Ma, Lingxiao and Yang, Fan and Mai, Luo},
  booktitle={19th USENIX Symposium on Operating Systems Design and Implementation (OSDI 25)},
  pages={257--273},
  year={2025}
}

@inproceedings{hwang2024pre,
  title={Pre-gated moe: An algorithm-system co-design for fast and scalable mixture-of-expert inference},
  author={Hwang, Ranggi and Wei, Jianyu and Cao, Shijie and Hwang, Changho and Tang, Xiaohu and Cao, Ting and Yang, Mao},
  booktitle={2024 ACM/IEEE 51st Annual International Symposium on Computer Architecture (ISCA)},
  pages={1018--1031},
  year={2024},
  organization={IEEE}
}

@inproceedings{he2025hydra,
  title={Hydra: Harnessing expert popularity for efficient mixture-of-expert inference on Chiplet system},
  author={He, Siqi and Zhu, Haozhe and Zheng, Jiapei and Wu, Lizhou and Jiao, Bo and Liu, Qi and Zeng, Xiaoyang and Chen, Chixiao},
  booktitle={2025 62nd ACM/IEEE Design Automation Conference (DAC)},
  pages={1--7},
  year={2025},
  organization={IEEE}
}

@article{ma20253d,
  title={3d-moe: A mixture-of-experts multi-modal llm for 3d vision and pose diffusion via rectified flow},
  author={Ma, Yueen and Zhuang, Yuzheng and Hao, Jianye and King, Irwin},
  journal={arXiv preprint arXiv:2501.16698},
  year={2025}
}

@inproceedings{kim2024monde,
  title={Monde: Mixture of near-data experts for large-scale sparse models},
  author={Kim, Taehyun and Choi, Kwanseok and Cho, Youngmock and Cho, Jaehoon and Lee, Hyuk-Jae and Sim, Jaewoong},
  booktitle={Proceedings of the 61st ACM/IEEE Design Automation Conference},
  pages={1--6},
  year={2024}
}

@article{wang2025d,
  title={D$^{2}$MoE: Dual Routing and Dynamic Scheduling for Efficient On-Device MoE-based LLM Serving},
  author={Wang, Haodong and Zhou, Qihua and Hong, Zicong and Guo, Song},
  journal={arXiv preprint arXiv:2504.15299},
  year={2025}
}

@article{eliseev2023fast,
  title={Fast inference of mixture-of-experts language models with offloading},
  author={Eliseev, Artyom and Mazur, Denis},
  journal={arXiv preprint arXiv:2312.17238},
  year={2023}
}

@article{tang2024hobbit,
  title={Hobbit: A mixed precision expert offloading system for fast moe inference},
  author={Tang, Peng and Liu, Jiacheng and Hou, Xiaofeng and Pu, Yifei and Wang, Jing and Heng, Pheng-Ann and Li, Chao and Guo, Minyi},
  journal={arXiv preprint arXiv:2411.01433},
  year={2024}
}

@inproceedings{fang2025klotski,
  title={Klotski: Efficient Mixture-of-Expert Inference via Expert-Aware Multi-Batch Pipeline},
  author={Fang, Zhiyuan and Huang, Yuegui and Hong, Zicong and Lyu, Yufeng and Chen, Wuhui and Yu, Yue and Yu, Fan and Zheng, Zibin},
  booktitle={Proceedings of the 30th ACM International Conference on Architectural Support for Programming Languages and Operating Systems, Volume 2},
  pages={574--588},
  year={2025}
}

@inproceedings{zhang2025daop,
  title={DAOP: Data-Aware Offloading and Predictive Pre-Calculation for Efficient MoE Inference},
  author={Zhang, Yujie and Aggarwal, Shivam and Mitra, Tulika},
  booktitle={2025 Design, Automation \& Test in Europe Conference (DATE)},
  pages={1--7},
  year={2025},
  organization={IEEE}
}

@inproceedings{cao2025moe,
  title={Moe-lightning: High-throughput moe inference on memory-constrained gpus},
  author={Cao, Shiyi and Liu, Shu and Griggs, Tyler and Schafhalter, Peter and Liu, Xiaoxuan and Sheng, Ying and Gonzalez, Joseph E and Zaharia, Matei and Stoica, Ion},
  booktitle={Proceedings of the 30th ACM International Conference on Architectural Support for Programming Languages and Operating Systems, Volume 1},
  pages={715--730},
  year={2025}
}

@inproceedings{ChipletDesignRule,
author = {Sangiovanni-Vincentelli, Alberto and Liang, Zheng and Zhou, Zhe and Zhang, Jiaxi},
title = {Automated Design of Chiplets},
year = {2023},
isbn = {9781450399784},
publisher = {Association for Computing Machinery},
address = {New York, NY, USA},
url = {https://doi.org/10.1145/3569052.3578917},
doi = {10.1145/3569052.3578917},
booktitle = {Proceedings of the 2023 International Symposium on Physical Design},
pages = {1–8},
numpages = {8},
location = {Virtual Event, USA},
series = {ISPD '23}
}

@article{zhu2025megascale,
  title={MegaScale-Infer: Serving Mixture-of-Experts at Scale with Disaggregated Expert Parallelism},
  author={Zhu, Ruidong and Jiang, Ziheng and Jin, Chao and Wu, Peng and Stuardo, Cesar A and Wang, Dongyang and Zhang, Xinlei and Zhou, Huaping and Wei, Haoran and Cheng, Yang and others},
  journal={arXiv preprint arXiv:2504.02263},
  year={2025}
}

@inproceedings{suo2025coserve,
  title={CoServe: Efficient Collaboration-of-Experts (CoE) Model Inference with Limited Memory},
  author={Suo, Jiashun and Liao, Xiaojian and Xiao, Limin and Ruan, Li and Wang, Jinquan and Su, Xiao and Huo, Zhisheng},
  booktitle={Proceedings of the 30th ACM International Conference on Architectural Support for Programming Languages and Operating Systems, Volume 2},
  pages={178--191},
  year={2025}
}

@article{he2025capacity,
  title={Capacity-Aware Inference: Mitigating the Straggler Effect in Mixture of Experts},
  author={He, Shwai and Cai, Weilin and Huang, Jiayi and Li, Ang},
  journal={arXiv preprint arXiv:2503.05066},
  year={2025}
}

@article{huang2025hd,
  title={HD-MoE: Hybrid and Dynamic Parallelism for Mixture-of-Expert LLMs with 3D Near-Memory Processing},
  author={Huang, Haochen and Zhong, Shuzhang and Zhang, Zhe and Li, Shuangchen and Niu, Dimin and Zheng, Hongzhong and Wang, Runsheng and Li, Meng},
  journal={arXiv preprint arXiv:2509.09420},
  year={2025}
}

@article{yu2025orders,
  title={Orders in Chaos: Enhancing Large-Scale MoE LLM Serving with Data Movement Forecasting},
  author={Yu, Zhongkai and Guan, Yue and Yu, Zihao and Zhou, Chenyang and Pei, Shuyi and Kang, Yangwook and Ding, Yufei and Tsai, Po-An},
  journal={arXiv preprint arXiv:2510.05497},
  year={2025}
}

@article{liu2024deepseek,
  title={Deepseek-v3 technical report},
  author={Liu, Aixin and Feng, Bei and Xue, Bing and Wang, Bingxuan and Wu, Bochao and Lu, Chengda and Zhao, Chenggang and Deng, Chengqi and Zhang, Chenyu and Ruan, Chong and others},
  journal={arXiv preprint arXiv:2412.19437},
  year={2024}
}

@article{shan2022architecture,
  title={Architecture of computing system based on chiplet},
  author={Shan, Guangbao and Zheng, Yanwen and Xing, Chaoyang and Chen, Dongdong and Li, Guoliang and Yang, Yintang},
  journal={Micromachines},
  volume={13},
  number={2},
  pages={205},
  year={2022},
  publisher={MDPI}
}

@article{zhang2023indm,
  title={INDM: Chiplet-based interconnect network and dataflow mapping for DNN accelerators},
  author={Zhang, Jinming and Fan, Xi and Ye, Yaoyao and Wang, Xuyan and Xiong, Guojie and Leng, Xianglun and Xu, Ningyi and Lian, Yong and He, Guanghui},
  journal={IEEE Transactions on Computer-Aided Design of Integrated Circuits and Systems},
  volume={43},
  number={4},
  pages={1107--1120},
  year={2023},
  publisher={IEEE}
}

@inproceedings{wang2024moa,
  title={Moa: Mixture-of-attention for subject-context disentanglement in personalized image generation},
  author={Wang, Kuan-Chieh and Ostashev, Daniil and Fang, Yuwei and Tulyakov, Sergey and Aberman, Kfir},
  booktitle={SIGGRAPH Asia 2024 Conference Papers},
  pages={1--12},
  year={2024}
}

@article{jin2024moh,
  title={Moh: Multi-head attention as mixture-of-head attention},
  author={Jin, Peng and Zhu, Bo and Yuan, Li and Yan, Shuicheng},
  journal={arXiv preprint arXiv:2410.11842},
  year={2024}
}

@article{shen2024jetmoe,
  title={Jetmoe: Reaching llama2 performance with 0.1 m dollars},
  author={Shen, Yikang and Guo, Zhen and Cai, Tianle and Qin, Zengyi},
  journal={arXiv preprint arXiv:2404.07413},
  year={2024}
}

@online{qwen1.5,
  title={Qwen1.5-MoE: Matching 7B Model Performance with 1/3 Activated Parameters},
  author={Qwen Team},
  url={https://qwenlm.github.io/blog/qwen-moe/},
  year={2024}
}

@article{wu2024yuan,
  title={Yuan 2.0-m32: Mixture of experts with attention router},
  author={Wu, Shaohua and Luo, Jiangang and Chen, Xi and Li, Lingjun and Zhao, Xudong and Yu, Tong and Wang, Chao and Wang, Yue and Wang, Fei and Qiao, Weixu and others},
  journal={arXiv preprint arXiv:2405.17976},
  year={2024}
}

@article{zhu2024llama,
  title={Llama-moe: Building mixture-of-experts from llama with continual pre-training},
  author={Zhu, Tong and Qu, Xiaoye and Dong, Daize and Ruan, Jiacheng and Tong, Jingqi and He, Conghui and Cheng, Yu},
  journal={arXiv preprint arXiv:2406.16554},
  year={2024}
}

@article{abouelenin2025phi,
  title={Phi-4-mini technical report: Compact yet powerful multimodal language models via mixture-of-loras},
  author={Abouelenin, Abdelrahman and Ashfaq, Atabak and Atkinson, Adam and Awadalla, Hany and Bach, Nguyen and Bao, Jianmin and Benhaim, Alon and Cai, Martin and Chaudhary, Vishrav and Chen, Congcong and others},
  journal={arXiv preprint arXiv:2503.01743},
  year={2025}
}

@article{liu2024grin,
  title={Grin: Gradient-informed moe},
  author={Liu, Liyuan and Kim, Young Jin and Wang, Shuohang and Liang, Chen and Shen, Yelong and Cheng, Hao and Liu, Xiaodong and Tanaka, Masahiro and Wu, Xiaoxia and Hu, Wenxiang and others},
  journal={arXiv preprint arXiv:2409.12136},
  year={2024}
}

@inproceedings{zhong2024adapmoe,
  title={AdapMoE: Adaptive sensitivity-based expert gating and management for efficient moe inference},
  author={Zhong, Shuzhang and Liang, Ling and Wang, Yuan and Wang, Runsheng and Huang, Ru and Li, Meng},
  booktitle={Proceedings of the 43rd IEEE/ACM International Conference on Computer-Aided Design},
  pages={1--9},
  year={2024}
}

@article{muennighoff2024olmoe,
  title={Olmoe: Open mixture-of-experts language models},
  author={Muennighoff, Niklas and Soldaini, Luca and Groeneveld, Dirk and Lo, Kyle and Morrison, Jacob and Min, Sewon and Shi, Weijia and Walsh, Pete and Tafjord, Oyvind and Lambert, Nathan and others},
  journal={arXiv preprint arXiv:2409.02060},
  year={2024}
}

@article{liu2025muon,
  title={Muon is scalable for LLM training},
  author={Liu, Jingyuan and Su, Jianlin and Yao, Xingcheng and Jiang, Zhejun and Lai, Guokun and Du, Yulun and Qin, Yidao and Xu, Weixin and Lu, Enzhe and Yan, Junjie and others},
  journal={arXiv preprint arXiv:2502.16982},
  year={2025}
}

@inproceedings{li2023accelerating,
  title={Accelerating distributed $\{$MoE$\}$ training and inference with lina},
  author={Li, Jiamin and Jiang, Yimin and Zhu, Yibo and Wang, Cong and Xu, Hong},
  booktitle={2023 USENIX Annual Technical Conference (USENIX ATC 23)},
  pages={945--959},
  year={2023}
}

@inproceedings{shi2023pipemoe,
  title={Pipemoe: Accelerating mixture-of-experts through adaptive pipelining},
  author={Shi, Shaohuai and Pan, Xinglin and Chu, Xiaowen and Li, Bo},
  booktitle={IEEE INFOCOM 2023-IEEE Conference on Computer Communications},
  pages={1--10},
  year={2023},
  organization={IEEE}
}

@inproceedings{punniyamurthy2024optimizing,
  title={Optimizing distributed ml communication with fused computation-collective operations},
  author={Punniyamurthy, Kishore and Hamidouche, Khaled and Beckmann, Bradford M},
  booktitle={SC24: International Conference for High Performance Computing, Networking, Storage and Analysis},
  pages={1--17},
  year={2024},
  organization={IEEE}
}

@article{go2025moetuner,
  title={Moetuner: Optimized mixture of expert serving with balanced expert placement and token routing},
  author={Go, Seokjin and Mahajan, Divya},
  journal={arXiv preprint arXiv:2502.06643},
  year={2025}
}

@inproceedings{rajbhandari2022deepspeed,
  title={Deepspeed-moe: Advancing mixture-of-experts inference and training to power next-generation ai scale},
  author={Rajbhandari, Samyam and Li, Conglong and Yao, Zhewei and Zhang, Minjia and Aminabadi, Reza Yazdani and Awan, Ammar Ahmad and Rasley, Jeff and He, Yuxiong},
  booktitle={International conference on machine learning},
  pages={18332--18346},
  year={2022},
  organization={PMLR}
}

@article{yu2024moesys,
  title={Moesys: A distributed and efficient mixture-of-experts training and inference system for internet services},
  author={Yu, Dianhai and Shen, Liang and Hao, Hongxiang and Gong, Weibao and Wu, Huachao and Bian, Jiang and Dai, Lirong and Xiong, Haoyi},
  journal={IEEE Transactions on Services Computing},
  volume={17},
  number={5},
  pages={2626--2639},
  year={2024},
  publisher={IEEE}
}

@article{doucet2025harmoeny,
  title={HarMoEny: Efficient Multi-GPU Inference of MoE Models},
  author={Doucet, Zachary and Sharma, Rishi and de Vos, Martijn and Pires, Rafael and Kermarrec, Anne-Marie and Balmau, Oana},
  journal={arXiv preprint arXiv:2506.12417},
  year={2025}
}

@article{zhao2023pytorch,
  title={Pytorch fsdp: experiences on scaling fully sharded data parallel},
  author={Zhao, Yanli and Gu, Andrew and Varma, Rohan and Luo, Liang and Huang, Chien-Chin and Xu, Min and Wright, Less and Shojanazeri, Hamid and Ott, Myle and Shleifer, Sam and others},
  journal={arXiv preprint arXiv:2304.11277},
  year={2023}
}

@inproceedings{wang2025harnessing,
  title={Harnessing inter-gpu shared memory for seamless moe communication-computation fusion},
  author={Wang, Hulin and Xia, Yaqi and Yang, Donglin and Zhou, Xiaobo and Cheng, Dazhao},
  booktitle={Proceedings of the 30th ACM SIGPLAN Annual Symposium on Principles and Practice of Parallel Programming},
  pages={170--182},
  year={2025}
}

@inproceedings{pan2025fsmoe,
  title={Fsmoe: A flexible and scalable training system for sparse mixture-of-experts models},
  author={Pan, Xinglin and Lin, Wenxiang and Zhang, Lin and Shi, Shaohuai and Tang, Zhenheng and Wang, Rui and Li, Bo and Chu, Xiaowen},
  booktitle={Proceedings of the 30th ACM International Conference on Architectural Support for Programming Languages and Operating Systems, Volume 1},
  pages={524--539},
  year={2025}
}

@article{du2024sida,
  title={Sida: Sparsity-inspired data-aware serving for efficient and scalable large mixture-of-experts models},
  author={Du, Zhixu and Li, Shiyu and Wu, Yuhao and Jiang, Xiangyu and Sun, Jingwei and Zheng, Qilin and Wu, Yongkai and Li, Ang and Li, Hai and Chen, Yiran},
  journal={Proceedings of Machine Learning and Systems},
  volume={6},
  pages={224--238},
  year={2024}
}

@inproceedings{yao2024exploiting,
  title={Exploiting inter-layer expert affinity for accelerating mixture-of-experts model inference},
  author={Yao, Jinghan and Anthony, Quentin and Shafi, Aamir and Subramoni, Hari and Panda, Dhabaleswar K DK},
  booktitle={2024 IEEE International Parallel and Distributed Processing Symposium (IPDPS)},
  pages={915--925},
  year={2024},
  organization={IEEE}
}

@article{song2024promoe,
  title={Promoe: Fast moe-based llm serving using proactive caching},
  author={Song, Xiaoniu and Zhong, Zihang and Chen, Rong and Chen, Haibo},
  journal={arXiv preprint arXiv:2410.22134},
  year={2024}
}

@article{li2025speculative,
  title={Speculative MoE: Communication Efficient Parallel MoE Inference with Speculative Token and Expert Pre-scheduling},
  author={Li, Yan and Zheng, Pengfei and Chen, Shuang and Xu, Zewei and Lai, Yuanhao and Du, Yunfei and Wang, Zhengang},
  journal={arXiv preprint arXiv:2503.04398},
  year={2025}
}

@article{he2024expertflow,
  title={Expertflow: Optimized expert activation and token allocation for efficient mixture-of-experts inference},
  author={He, Xin and Zhang, Shunkang and Wang, Yuxin and Yin, Haiyan and Zeng, Zihao and Shi, Shaohuai and Tang, Zhenheng and Chu, Xiaowen and Tsang, Ivor and Soon, Ong Yew},
  journal={arXiv preprint arXiv:2410.17954},
  year={2024}
}

@article{kossmann2022optimizing,
  title={Optimizing mixture of experts using dynamic recompilations},
  author={Kossmann, Ferdinand and Jia, Zhihao and Aiken, Alex},
  journal={arXiv preprint arXiv:2205.01848},
  year={2022}
}

@article{li2025static,
  title={Static Batching of Irregular Workloads on GPUs: Framework and Application to Efficient MoE Model Inference},
  author={Li, Yinghan and Li, Yifei and Zhang, Jiejing and Chen, Bujiao and Chen, Xiaotong and Duan, Lian and Jin, Yejun and Li, Zheng and Liu, Xuanyu and Wang, Haoyu and others},
  journal={arXiv preprint arXiv:2501.16103},
  year={2025}
}

@article{li2021sequence,
  title={Sequence parallelism: Long sequence training from system perspective},
  author={Li, Shenggui and Xue, Fuzhao and Baranwal, Chaitanya and Li, Yongbin and You, Yang},
  journal={arXiv preprint arXiv:2105.13120},
  year={2021}
}

@article{merity2016pointer,
  title={Pointer sentinel mixture models},
  author={Merity, Stephen and Xiong, Caiming and Bradbury, James and Socher, Richard},
  journal={arXiv preprint arXiv:1609.07843},
  year={2016}
}

@article{raffel2020exploring,
  title={Exploring the limits of transfer learning with a unified text-to-text transformer},
  author={Raffel, Colin and Shazeer, Noam and Roberts, Adam and Lee, Katherine and Narang, Sharan and Matena, Michael and Zhou, Yanqi and Li, Wei and Liu, Peter J},
  journal={Journal of machine learning research},
  volume={21},
  number={140},
  pages={1--67},
  year={2020}
}

@article{sakaguchi2021winogrande,
  title={Winogrande: An adversarial winograd schema challenge at scale},
  author={Sakaguchi, Keisuke and Bras, Ronan Le and Bhagavatula, Chandra and Choi, Yejin},
  journal={Communications of the ACM},
  volume={64},
  number={9},
  pages={99--106},
  year={2021},
  publisher={ACM New York, NY, USA}
}

@article{agrawal2023sarathi,
  title={Sarathi: Efficient llm inference by piggybacking decodes with chunked prefills},
  author={Agrawal, Amey and Panwar, Ashish and Mohan, Jayashree and Kwatra, Nipun and Gulavani, Bhargav S and Ramjee, Ramachandran},
  journal={arXiv preprint arXiv:2308.16369},
  year={2023}
}

@inproceedings{chou2022netflex,
  title={NetFlex: A 22nm multi-chiplet perception accelerator in high-density fan-out wafer-level packaging},
  author={Chou, Teyuh and Tang, Wei and Rotaru, Mihai D and Liu, Chester and Dutta, Rahul and Siang, Sharon Lim Pei and Wee, David Ho Soon and Bhattacharya, Surya and Zhang, Zhengya},
  booktitle={2022 IEEE Symposium on VLSI Technology and Circuits (VLSI Technology and Circuits)},
  pages={208--209},
  year={2022},
  organization={IEEE}
}

@inproceedings{tu202316,
  title={16.4 TensorCIM: A 28nm 3.7 nJ/gather and 8.3 TFLOPS/W FP32 digital-CIM tensor processor for MCM-CIM-based beyond-NN acceleration},
  author={Tu, Fengbin and Wang, Yiqi and Wu, Zihan and Wu, Weiwei and Liu, Leibo and Hu, Yang and Wei, Shaojun and Yin, Shouyi},
  booktitle={2023 IEEE International Solid-State Circuits Conference (ISSCC)},
  pages={254--256},
  year={2023},
  organization={IEEE}
}

@inproceedings{zhu2022comb,
  title={COMB-MCM: Computing-on-memory-boundary NN processor with bipolar bitwise sparsity optimization for scalable multi-chiplet-module edge machine learning},
  author={Zhu, Haozhe and Jiao, Bo and Zhang, Jinshan and Jia, Xinru and Wang, Yunzhengmao and Guan, Tianchan and Wang, Shengcheng and Niu, Dimin and Zheng, Hongzhong and Chen, Chixiao and others},
  booktitle={2022 IEEE International Solid-State Circuits Conference (ISSCC)},
  volume={65},
  pages={1--3},
  year={2022},
  organization={IEEE}
}

@inproceedings{tan2023scalable,
  title={A scalable multi-chiplet deep learning accelerator with hub-side 2.5 D heterogeneous integration},
  author={Tan, Zhanhong and Wu, Yifu and Zhang, Yannian and Shi, Haobing and Zhang, Wuke and Ma, Kaisheng},
  booktitle={2023 IEEE Hot Chips 35 Symposium (HCS)},
  pages={1--17},
  year={2023},
  organization={IEEE}
}

@article{sharma2022universal,
  title={Universal chiplet interconnect express (UCIe): An open industry standard for innovations with chiplets at package level},
  author={Sharma, Debendra Das and Pasdast, Gerald and Qian, Zhiguo and Aygun, Kemal},
  journal={IEEE Transactions on Components, Packaging and Manufacturing Technology},
  volume={12},
  number={9},
  pages={1423--1431},
  year={2022},
  publisher={IEEE}
}

@inproceedings{mota2023ucie,
  author    = {Mota, G.},
  title     = {{UCIe: Universal Chiplet Interconnect Express}},
  booktitle = {Chiplet Summit},
  year      = {2023},
  month     = {jan},
  note      = {[Online]. Available: \url{https://chipletsummit.com/proceeding_files/a0q5f000001WuE0/20230126_A-201_Mota.PDF}}
}

@inproceedings{lin202536,
  title={36.1 A 32Gb/s 10.5 Tb/s/mm 0.6 pJ/b UCIe-Compliant Low-Latency Interface in 3nm Featuring Matched-Delay for Dynamic Clock Gating},
  author={Lin, Mu-Shan and Tsai, Chien-Chun and Li, Shenggao and Chen, Wei-Chih and Huang, Wen-Hung and Chen, Yu-Chi and Huang, Yu-Jie and Drake, Alan and Wen, Chin-Hua and Ranucci, Paul and others},
  booktitle={2025 IEEE International Solid-State Circuits Conference (ISSCC)},
  volume={68},
  pages={586--588},
  year={2025},
  organization={IEEE}
}

@inproceedings{melek20250,
  title={A 0.29 pJ/b 5.27 Tb/s/mm UCIe Advanced Package Link in 3nm FinFET with 2.5 D CoWoS Packaging},
  author={Melek, Didem Turker and Navinkumar, R and Vandersand, James and Sarkar, Pyare and Prakash, BS and Leuciuc, Adrian and Geary, Kevin and Ma, Shaojun and Mehta, Chirag Mukesh and Jain, Shashi and others},
  booktitle={2025 IEEE International Solid-State Circuits Conference (ISSCC)},
  volume={68},
  pages={590--592},
  year={2025},
  organization={IEEE}
}

@inproceedings{vandersand20250,
  title={A 0.52 pJ/bit 0.448 Tbps/mm UCIe Standard Package Die-to-Die Transceiver with Low-Latency TX Clock Alignment in 3nm FinFET},
  author={Vandersand, James and Melek, Didem Turker and Geary, Kevin and BS, Prakash and Jain, Shashi and Bothra, Basant and Sarkar, Pyare and Sabharwal, Pawan and Navinkumar, R and Chang, Ken},
  booktitle={2025 Symposium on VLSI Technology and Circuits (VLSI Technology and Circuits)},
  pages={1--3},
  year={2025},
  organization={IEEE}
}

@inproceedings{jiao202537,
  title={37.4 SHINSAI: A 586mm 2 Reusable Active TSV Interposer with Programmable Interconnect Fabric and 512Mb 3D Underdeck Memory},
  author={Jiao, Bo and Zhu, Haozhe and Zeng, Yuman and Li, Yongjiang and Liao, Jie and Jia, Siyao and Tian, Mochen and Chen, Zexing and Zhu, Jundong and Wen, Dexin and others},
  booktitle={2025 IEEE International Solid-State Circuits Conference (ISSCC)},
  volume={68},
  pages={01--03},
  year={2025},
  organization={IEEE}
}

@article{team2024phi,
  title={Phi-3 technical report: A highly capable language model locally on your phone},
  author={Team, Phi and others},
  journal={arXiv preprint arXiv:2404.14219},
  year={2024}
}

@article{qu2025mldse,
  title={MLDSE: Scaling Design Space Exploration Infrastructure for Multi-Level Hardware},
  author={Qu, Huanyu and Zhang, Weihao and Lin, Junfeng and Ma, Songchen and Li, Hongyi and Shi, Luping and Xu, Chengzhong},
  journal={arXiv preprint arXiv:2503.21297},
  year={2025}
}


\end{document}